\documentclass{emulateapj}
\usepackage{ulem}

\slugcomment{draft v2}

\shorttitle{Energy Sources and Light Curves of Macronovae}
\shortauthors{Kisaka, Ioka \& Takami}

\begin{document}


\title{Energy Sources and Light curves of Macronovae}


\author{Shota Kisaka\altaffilmark{1}}
\email{kisaka@post.kek.jp}
\author{Kunihito Ioka\altaffilmark{1,2}}
\email{kunihito.ioka@kek.jp}
\author{Hajime Takami\altaffilmark{1,3}}
\email{takami@post.kek.jp}


\altaffiltext{1}{Theory Center, Institute of Particle and Nuclear Studies, KEK, Tsukuba 305-0801, Japan}
\altaffiltext{2}{Department of Particle and Nuclear Physics, the Graduate University for Advanced Studies (Sokendai), Tsukuba 305-0801, Japan}
\altaffiltext{3}{JSPS Research Fellow}


\begin{abstract}
A macronova (kilonova) was discovered with a short gamma-ray burst, GRB 130603B, which is widely believed to be powered by the radioactivity of $r$-process elements synthesized in the ejecta of a neutron star binary merger. As an alternative, we propose that macronovae are energized by the central engine, i.e., a black hole or neutron star, and the injected energy is emitted after the adiabatic expansion of ejecta. This engine model is motivated by extended emission of short GRBs. In order to compare the theoretical models with observations, we develop analytical formulae for the light curves of macronovae. The engine model allows a wider parameter range, especially smaller ejecta mass, and better fit to observations than the $r$-process model. Future observations of electromagnetic counterparts of gravitational waves should distinguish energy sources and constrain the activity of central engine and the $r$-process nucleosynthesis.
\end{abstract}





\section{INTRODUCTION}
\label{interduction}

 \begin{figure*}
  \begin{center}
   \includegraphics[width=150mm]{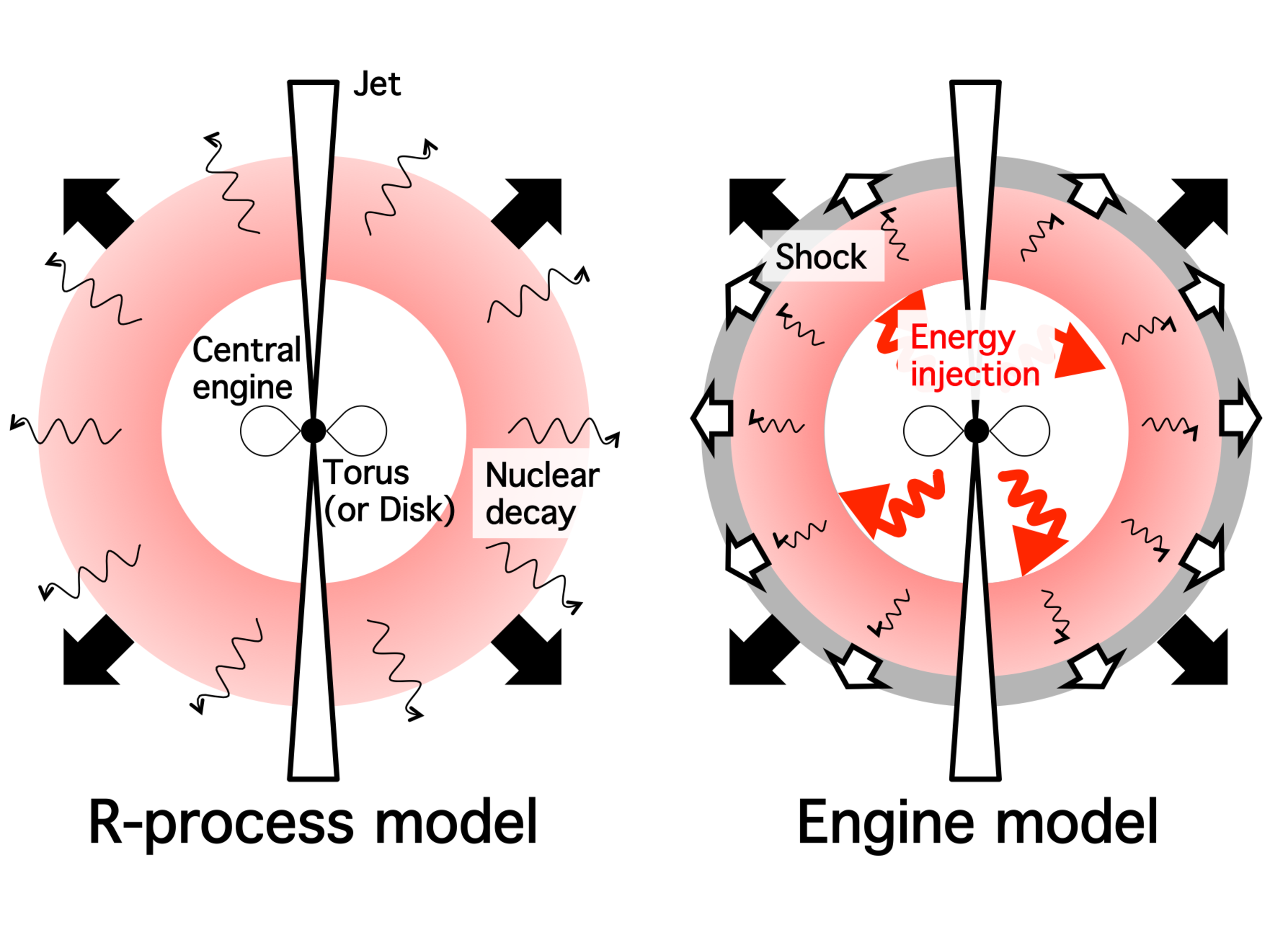}
   \caption{Schematic pictures of the $r$-process model (left) and the engine model (right).}
   \label{figure:model}
  \end{center}
 \end{figure*}

Gravitational wave (GW) observations are expected to provide a new view of relativistic phenomena in the Universe. One of the most promising candidates for the direct detection of GWs is the merger of compact binaries such as binary neutron stars (NSs). The second generation of ground-based GW detectors, such as Advanced LIGO \citep{Aba+10}, Advanced VIRGO \citep{Ace+14} and KAGRA \citep{Kur+10}, will reach the sensitivity required to detect GWs from the inspiral and coalescence of compact binary systems including binary NSs within a few hundred Mpc. Statistical studies suggest that a few tens of merger events should be observed per year \citep{Aba+10b}.

Electromagnetic counterparts of GW emitters have been recently focused on to maximize a scientific return from the expected detection of GWs \citep[e.g., ][]{MB12}. Follow-up observations of these electromagnetic counterparts are important to confirm a GW detection and to investigate progenitors and environments. The electromagnetic detection also improves the localization of GW sources because the localization accuracy by photons is much better than that by the ground-based GW detectors $\sim10-100$ deg$^2$ \citep[e.g., ][]{Ess+14}.

Sophisticated simulations have revealed mass ejection associated with the mergers of binary NSs by several mechanisms. Significant mass is dynamically ejected by gravitational torques and hydrodynamical interactions during the mergers, called dynamical ejecta \citep[e.g., ][]{Ros+99, RJ01, Hot+13}. General relativistic simulations show that these ejecta distribute nearly isotropic compared to Newtonian simulations in the cases of binary NSs \citep{Hot+13}, while they are anisotropic for NS-black hole (BH) mergers \citep{KIS13}. Mass may be also ejected through winds driven by neutrinos \citep{Des+09}, magnetic fields of and/or amplified by the merged objects \citep{SSKI11, KKS12, KKSSW14}, viscous heating and nuclear recombination \citep{FM13, FKMQ14}. 

A traditional electromagnetic counterpart is short-hard gamma-ray bursts \citep[GRBs; ][]{NPP92}. Recent simulations have revealed that a hypermassive NS is formed from the merger of a NS binary \citep[e.g., ][]{Hot+13}, which is believed to collapse into a BH at later time. Non-collapsed matter and some ejecta falling back to the BH form a torus around the BH \citep[e.g., ][]{Ros07}. Then, a relativistic jet may be launched from the BH-torus system, which is believed to be the central engine of short-hard GRBs. Another interesting possibility is a so-called macronova/kilonova, which is thermal emission from ejecta \citep[e.g., ][]{LP98, Kul05, BK13}. The radiative energy of a macronova is estimated between that of a classical nova and supernova. Ejecta can also produce non-thermal emission at later time similarly to supernova remnants \citep{NP11, PNR13, TKI14}. Ejecta may accompany an advanced relativistic part, producing early emission \citep[$\sim$ hours; ][]{KIS14, Met+14}. Emission from macronovae and NS binary merger remnants is almost isotropic and hence different from that of short GRBs which depends on the directions of their relativistic jets. Moreover, macronovae are closer in time to mergers than emission from merger remnants and do not depend on the properties of circumburst environments. Therefore, macronovae are expected to play a crucial role to localize a large sample of GW events \citep{MB12}. 

Recently, a macronova candidate following GRB 130603B was discovered \citep{Tan+13, BFC13}. This candidate is widely interpreted as the results of the radioactive decay of $r$-process elements produced in the ejecta of a compact binary merger \citep{Tan+13, BFC13, Hot+13b, PKR14, Gro+14}. We call this scenario {\it an $r$-process model} throughout this paper. The ejecta from a merger of binary NSs is primarily neutron-rich. Then, heavy radioactive elements (mass number $\gtrsim 130$) are expected to form through neutron-capture onto nuclei ($r$-process nucleosynthesis) \citep[e.g., ][]{LS74}. Although the $r$-process nucleosynthesis ends a few hundred millisecond after a merger, synthesized elements release energy due to nuclear fission and beta decays up to $\sim100$ days \citep[e.g., ][]{Wan+14}. A schematic picture for this model is shown in the left panel of figure \ref{figure:model}. If this scenario is correct, the observations also give important insights into the enrichment of $r$-process elements in the galaxy evolution \citep[e.g., ][]{PKR14}. Although the $r$-process model explains the observed light curve of the macronova, it is based on the limited observational data and the nuclear heating rate with large uncertainties. Required mass of dynamical ejecta to explain the observations is relatively large compared with the simulation results \citep{Gro+14}. In addition, the occurrence of $r$-process nucleosynthesis needs the ejecta with low electron fraction ($Y_e\lesssim0.1$). However, relatively high electron fraction ($Y_e\sim0.2-0.5$) can be also realized, which has been discussed for neutrino-driven wind \citep[e.g., ][]{FM13}. It is worth considering other possibilities such as the scenarios of an external shock between ejecta and surrounding medium \citep{Jin+13}, a supramassive magnetar \citep{Fan+13} and dust grains \citep{TNI14}.

In this study, we consider another power source of macronovae, i.e., energy injection from the activity of the central engine, in addition to the radioactive decay of $r$-process elements. This is similar to the early evolution of core-collapse supernovae \citep[e.g., ][]{A80, P93}. We call this model {\it an engine model} throughout this paper. There are several motivations to consider that the activity of the central engine contributes to the heating of ejecta. One observational motivation is the extended emission following the prompt emission of short GRBs. The origin of extended emission is considered to be the activity of the central engine \citep{Bar+05} because the sharp drop of its light curve is difficult to be reproduced by afterglow emission \citep{IKZ05}. After the merger, a stable NS or a BH is formed. In the case that a BH with a torus (or disk) is formed, the energy injection to the ejecta is expected as a form of the jet and/or disk wind \citep[e.g., ][]{Nak+13}. In the case that a NS with strong poloidal magnetic field is formed as a result of a merger, the wind of relativistic particles is ejected \citep{Dai+06, MQT08, YZG13, WD13, MP14}. Then, the wind collides with the ejecta, and about half of the wind energy converts to the internal energy by the shock-heating. A schematic picture is shown in the right-hand side of figure \ref{figure:model}. 

The ejecta emission powered by a stable magnetar has already been discussed \citep{YZG13, WD13, MP14}. They suggest that the magnetar-powered ejecta emit the brighter optical and X-ray emissions than that of the $r$-process model. However, they did not show that the magnetar-powered ejecta explain the detected infrared excess in GRB 130603B \citep{Tan+13, BFC13}. 

The engine model can provide energy enough to reproduce the detected macronova candidate, GRB 130603B. We do not specify the specific heating sources. Alternatively, to estimate the luminosity and temperature, we assume that the internal energy $E_{\rm int0}\sim10^{51}$ erg is injected to the ejecta at the time $t_{\rm inj}\sim10^2$ s after the merger. These values are consistent with typical isotropic energy $E_{\rm iso}\sim10^{50}-10^{51}$ erg and duration $t_{\rm dur}\sim10-10^2$ s of the extended emission \citep{Sak+11}. Using the velocity of the ejecta $v$, the temperature at $t_{\rm inj}$ is $T_0\sim[E_{\rm int0}/(av^3t_{\rm inj}^3)]^{1/4}$, where $a$ is the radiative constant. If we only consider the adiabatic cooling for the cooling process of the ejecta, the evolution of the internal energy $E_{\rm int}$ and temperature $T$ is scaled as $E_{\rm int}\propto t^{-1}$ and $T\propto t^{-1}$. The luminosity is described as $L\sim E_{\rm int}/t$. Adopting the ejecta velocity $v\sim10^{10}$cm s$^{-1}$ \citep{Hot+13}, the luminosity $L$ and the temperature $T$ at $t\sim10^6$s are 
\begin{eqnarray}\label{sec1:L}
L &\sim& \frac{E_{\rm int0}}{t}\left(\frac{t}{t_{\rm inj}}\right)^{-1} \nonumber \\
&\sim&10^{41}\left(\frac{E_{\rm int0}}{10^{51}{\rm erg}}\right)\left(\frac{t_{\rm inj}}{10^2{\rm s}}\right)\left(\frac{t}{10^6{\rm s}}\right)^{-2}~{\rm erg}~{\rm s}^{-1},
\end{eqnarray}
and
\begin{eqnarray}\label{sec1:T}
T&\sim& T_0\left(\frac{t}{t_{\rm inj}}\right)^{-1} \nonumber \\
&\sim&2\times10^3\left(\frac{E_{\rm int0}}{10^{51}{\rm erg}}\right)^{1/4}\left(\frac{t_{\rm inj}}{10^2{\rm s}}\right)^{1/4}\nonumber \\
& &\times\left(\frac{v}{10^{10}{\rm cm}~{\rm s}^{-1}}\right)^{-3/4}\left(\frac{t}{10^6{\rm s}}\right)^{-1}~{\rm K}.
\end{eqnarray}
The observations of macronova of GRB 130603B give J-band luminosity $\sim10^{41}$ erg s$^{-1}$ and the difference between J-band and B-band $\gtrsim2.5$ mag which corresponds to the temperature $\lesssim4\times10^3$ K at $t\sim7$ days after GRB 130603B in the source rest frame \citep{Tan+13, BFC13}. Therefore, in this estimate, the luminosity and temperature for the engine model is consistent with the observation of the macronova following GRB130603B. 

We model the evolution of luminosity and temperature of a macronova. Unlike the previous studies, we treat the model in an analytical manner and formulate a light curve including the early phase ($\sim10^3-10^5$ s), which is important for the search of electromagnetic counterparts of GW emitters. We consider shock-heating due to the activity of a central engine as a heating mechanism of the ejecta. For comparison, the $r$-process model, which has been discussed in most papers \citep[e.g., ][]{LP98}, is also formulated. Then, we compare the results of our models with observations to constrain the model parameters such as the ejected mass and the velocity of the ejecta. Although our models are simplified, it is valuable to make comparison between two heating models. In section \ref{model}, we introduce our model assumptions. We describe the analytical models for the evolution of luminosity and temperature in section \ref{evolution}. Then, we compare our results with observations in section \ref{discuss}. Implications for the discrimination between two models are also discussed. We summarize our results in section \ref{summary}. In appendix A, we summarize the formulae for the observed temperature and bolometric luminosity.

\section{MODEL}
\label{model}

Significant mass of material $\sim10^{-3}-10^{-1}M_{\odot}$ is ejected during a binary merger. We model ejecta by following the results of the general relativistic simulations of NS-NS mergers in \citet{Hot+13}. The simulations show that ejecta expand in a nearly homologous manner \citep[see also ][]{Ros+14}. The morphology of the ejecta is quasi-spherical in the case of a merger of binary NSs. According to these results, we assume an isotropic and homologous expansion for the ejecta. Then, the velocity of ejecta $v$ is
\begin{eqnarray}\label{sec2:v}
v\sim r/t
\end{eqnarray}
where the radius $r$ originates the central engine and the time $t$ is measured from the time when a compact binary merges.

Note that in the case of a merger of NS-BH binary, the ejected mass expands with significant anisotropy \citep{KOST11, Fou+13, Fou+14, Lov+13, Dea+13, KIS13}. We do not consider such anisotropic ejecta in this work. 

 \begin{figure*}
  \begin{center}
   \includegraphics[width=150mm]{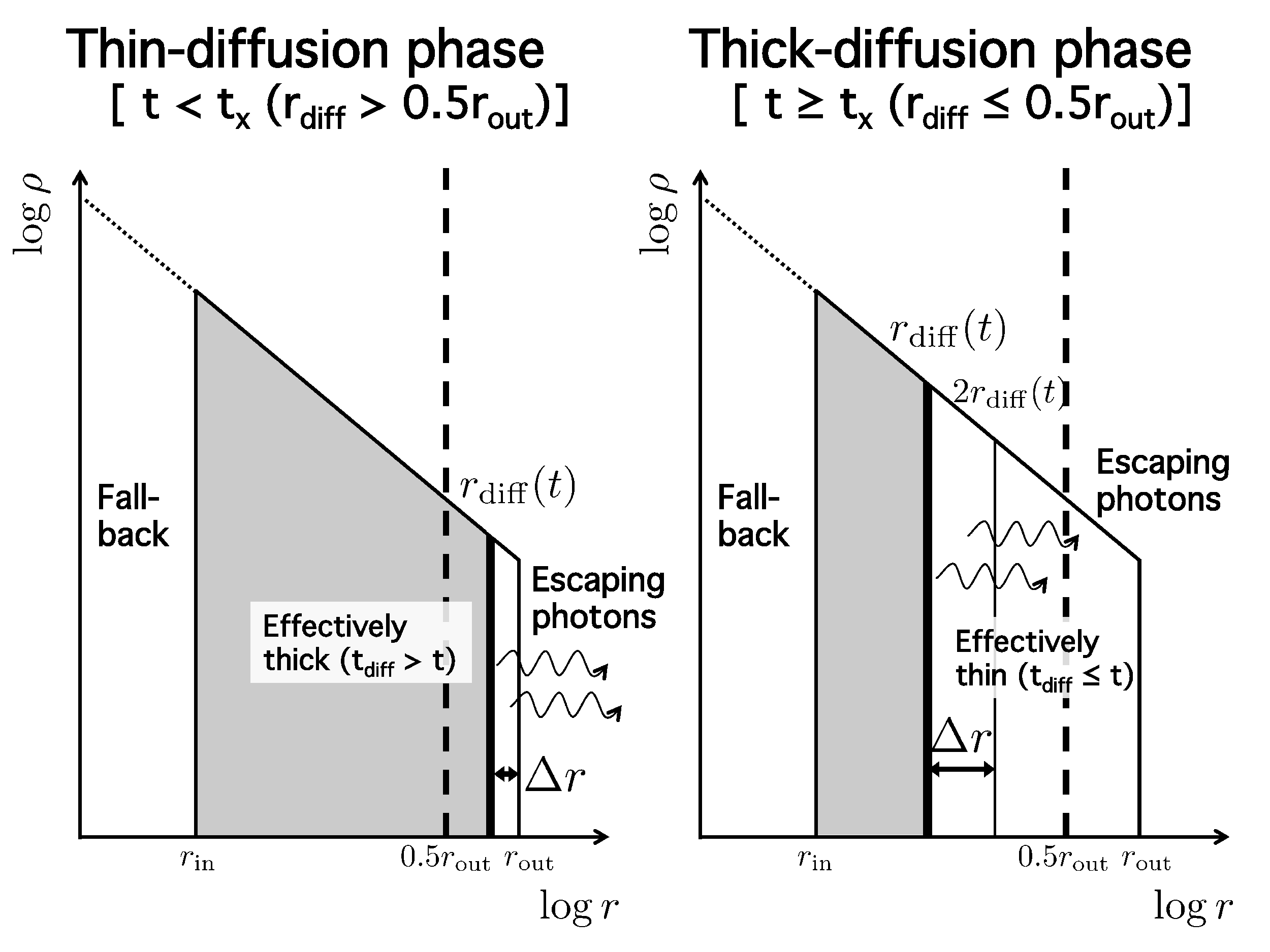}
   \caption{Schematic pictures of the thin- (left) and thick-diffusion phases (right). The horizontal and vertical axes show the radius which originates from the central engine and the mass density of the ejecta, respectively, in logarithmic scales. The radii $r_{\rm out}$ and $r_{\rm in}$ correspond to the outer and inner edges of the ejecta. Ejecta expand in a homologous manner ($v\sim r/t$). Material at the inner region up to the inner edge of the ejecta (dotted line) falls back to the central engine. The thick vertical lines show the diffusion radius $r_{\rm diff}$ at which the diffusion time equals to the dynamical time. Photons emitted from the right side of the thick vertical line (effectively thin region) can diffuse out from the ejecta. The effectively thick region corresponds to the shaded area. The size $\Delta r$ is the propagation distance to evaluate the diffusion time. Since most scatterings occur near the diffusion radius, we divide two phases whether the diffusion radius is larger than $0.5r_{\rm out}$ (thick vertical dashed lines) or not. The time $t_{\times}$ corresponds to the time when the diffusion radius equals to the half of the radius of the outer edge of the ejecta. See text for details.}
   \label{figure:diffuse}
  \end{center}
 \end{figure*}

\subsection{Density Profile}
\label{density}

\citet{Nag+14} found that the profile of ejecta obtained from simulations by \citet{Hot+13} can be well fitted by a power-law function $\rho\propto v^{-\beta}$. The power-law index of snapshot density $\beta$ is more or less independent on the dynamics of mergers, which is in the range of $\beta \sim 3$--4 for $v_{\min}\le v\le v_{\max}$, where $v_{\max}$ and $v_{\rm min}$ are the velocities of the outer and inner edges of the ejecta, respectively. We choose the middle of this range $\beta=3.5$ in this study. We also fix the maximum velocity $v_{\max}=0.4c$ from simulation results \citep{Hot+13}. The maximum velocity $v_{\max}$ is comparable with the escape velocity of the system. The minimum velocity $v_{\min}$ is mainly determined by complicated dynamics at the initial stage of the merger $t\ll10^2{\rm s}$. For the mass density profile at the front of the ejecta, we assume the discrete boundary and the mass density $\rho=0$ at the region $r>v_{\max}t$. Although this profile may be far from the actual one \footnote{The outer ejecta may have a relativistic component \citep{KIS14} and/or an exponential or a power-law profile (see Sec. \ref{outer}). Such a profile is difficult to calculate precisely with current numerical calculations.}, our main aim is to compare two models for energy sources, so that our conclusions are not affected. In section \ref{outer}, we discuss the dependence on the mass density profile at the outer region of the ejecta for the observed light curve. Here, we only consider the evolution after the initial stage of the merger ($t\gg t_{\rm inj}$) and treat $v_{\min}$ as a model parameter. Because of homologous expansion, the density decreases as $\rho \propto t^{-3}$. Then, the density profile is described by
\begin{eqnarray}\label{sec2-1:rho}
\rho (t,v)=\rho_0 \left(\frac{t}{t_0}\right)^{-3}\left(\frac{v}{v_{\min}}\right)^{-\beta}.
\end{eqnarray}
where $\rho_0$ and $t_0$ are normalization factors. The factor $\rho_0t_0^3$ is related to the total mass of the ejecta $M_{\rm ej}$ as following,
\begin{eqnarray}\label{sec2-1:M_ej}
M_{\rm ej}&=&4\pi\int_{v_{\min}t_0}^{v_{\max}t_0}\rho(t_0,v)r^2dr \nonumber \\
&=&\frac{4\pi}{\beta-3}\rho_0(v_{\min}t_0)^3\left[1-\left(\frac{v_{\max}}{v_{\min}}\right)^{3-\beta}\right],
\end{eqnarray}
where we use $dr(t=t_0)=t_0dv$ from equation (\ref{sec2:v}). We also introduce the radius of ejecta outer edge 
\begin{eqnarray}\label{sec2-1:r_out}
r_{\rm out}=v_{\max}t,
\end{eqnarray} 
and their inner edge
\begin{eqnarray}\label{sec2-1:r_in}
r_{\rm in}=v_{\min}t.
\end{eqnarray}

\subsection{Diffusion Radius}
\label{diffusion}

The inner part of the ejecta is optically thick, and therefore the propagation of radiation in the ejecta can be regarded as a diffusion process. Photons can diffusively escape from the region which satisfies that the diffusion time, $t_{\rm diff}$, is smaller than the dynamical time $t$,
\begin{eqnarray}\label{sec2-1:diffuse}
t_{\rm diff}\le t.
\end{eqnarray}
The medium in this region is called to be effectively thin \citep{RL79}. For convenience, we introduce a diffusion radius $r_{\rm diff}(t)$ which is the radius satisfying the condition $t=t_{\rm diff}$. Furthermore, we divide ejecta into two regions called the effectively thin ($r\ge r_{\rm diff}$) and effectively thick ($r < r_{\rm diff}$) regions. Near the diffusion radius, the optical depth is $\tau\gg1$. We consider random walk for photons so that the mean number of scatterings to propagate for the distance $\Delta r$ is $(\Delta r/l_{\rm mfp})^2$, where $l_{\rm mfp}$ is the mean free path for a photon. Hence, the diffusion time $t_{\rm diff}$ for the propagation distance $\Delta r$ is
\begin{eqnarray}\label{sec2-1:t_diff}
t_{\rm diff}\sim\frac{l_{\rm mfp}}{c}\left(\frac{\Delta r}{l_{\rm mfp}}\right)^2\sim\tau\frac{\Delta r}{c}.
\end{eqnarray}
In the right hand of equation (\ref{sec2-1:t_diff}), we use $\tau\sim\Delta r/l_{\rm mfp}$. 

We calculate the diffusion radius $r_{\rm diff}$ from the condition $t_{\rm diff}=t$. Since the mass density profile of the ejecta is described by a decreasing power-law function (equation \ref{sec2-1:rho}), the diffusion time $t_{\rm diff}$ is negligible in an outer part. Thus, in order to calculate the diffusion radius $r_{\rm diff}$, it is a good approximation to only consider scatterings near $r_{\rm diff}$ ($\Delta r\sim r_{\rm diff}$). However, in the early phase, the distance from the outer edge of the ejecta $r_{\rm out}$ to the diffusion radius $r_{\rm diff}$ is smaller than the diffusion radius $r_{\rm out}-r_{\rm diff}<r_{\rm diff}$. Therefore, we should take the propagation distance as 
\begin{eqnarray}\label{sec2-1:Deltar}
\Delta r\sim \left\{ \begin{array}{ll}
 r_{\rm out}-r_{\rm diff} &~~(r_{\rm diff}>0.5r_{\rm out}) \\
 & \\
 r_{\rm diff} &~~(r_{\rm diff}\le 0.5r_{\rm out}). \\
\end{array} \right.
\end{eqnarray}
We call the first the thin-diffusion phase and the second the thick-diffusion phase throughout this paper. We schematically show these two phases in figure \ref{figure:diffuse}. Note that in the thin-diffusion phase, since the size of the effectively thin region is much smaller than the size of the ejecta ($r_{\rm out}-r_{\rm in}$), the calculation of the radiative transfer using Monte Carlo technique \citep[e.g., ][]{BK13, TH13} requires a large number of realizations to follow the temporal evolution, which do not seem to have been considered properly so far. 

To obtain the diffusion radius, we need to calculate the optical depth $\tau$ of photons which propagate a distance $\Delta r$. Using equations (\ref{sec2-1:rho}) and (\ref{sec2-1:Deltar}), the optical depth $\tau$ is described as,
\begin{eqnarray}\label{sec2-1:tau}
\tau&=&\int_{r_{\rm diff}}^{r_{\rm out}}\kappa\rho dr \nonumber \\
&=&\frac{(\beta-3)\kappa M_{\rm ej}}{4\pi(\beta-1)v_{\min}^2t^2}\left[1-\left(\frac{v_{\max}}{v_{\min}}\right)^{3-\beta}\right]^{-1} \nonumber \\
& &\times
 {\displaystyle \left[\left(\frac{r_{\rm diff}}{v_{\min}t}\right)^{1-\beta}-\left(\frac{v_{\max}}{v_{\min}}\right)^{1-\beta}\right]},
\end{eqnarray}
in the thin-diffusion phase, and 
\begin{eqnarray}\label{sec2-1:tau2}
\tau&=&\int_{r_{\rm diff}}^{2r_{\rm diff}}\kappa\rho dr \nonumber \\
&=&\frac{(\beta-3)\kappa M_{\rm ej}}{4\pi(\beta-1)v_{\min}^2t^2}\left[1-\left(\frac{v_{\max}}{v_{\min}}\right)^{3-\beta}\right]^{-1} \nonumber \\
& &\times
 {\displaystyle \left(\frac{r_{\rm diff}}{v_{\min}t}\right)^{1-\beta}(1-2^{1-\beta})},
\end{eqnarray}
in the thick-diffusion phase, where $\kappa$ is the opacity of the ejecta. For simplicity, we use a grey approximation and a spatially uniform value of the opacity $\kappa$. From the results of \citet{TH13} \citep[see also ][]{KBB13} which consider the contribution from all $r$-process elements to the opacity of merger ejecta, the evolution of the bolometric luminosity can be approximately described by the constant value of the opacity, $\kappa\sim3-30$ cm$^2$g$^{-1}$. Following their results, we use this value for the opacity of the ejecta. Note that the exact value of the opacity of the ejecta has some uncertainties in the production efficiency of $r$-process elements and its spatial distribution. Moreover, if the ejecta temperature is low enough for dust formation ($T\lesssim2000$ K), the opacity significantly increases \citep{TNI14}. From these reasons, we consider the dependence on $\kappa$ in section \ref{evolution}.  

Our model is based on the formulation of the light curves of supernovae \citep[e.g., ][]{C92, NS10, RW11}, but there are several differences. In the case of type II supernovae, the opacity is significantly reduced due to hydrogen recombination \citep[e.g., ][]{GNS14}. However, since the ionization potentials of the lanthanides included in the $r$-process elements are generally lower than that of hydrogen and the iron group, the opacity remains high at relatively low temperature \citep{KBB13}. Therefore, we do not consider the recombination effects for the opacity.

As far as we know, the supernova studies \citep[e.g., ][]{C92, NS10, RW11} have not taken into account the thin-diffusion phase, which is necessary for treating the thickness of the diffusion length appropriately and estimating the physical quantities by the values at the outer edge of the ejecta in the analytical formulae. This phase may be also important for the case of supernovae.

Some supernova studies consider the planar phase \citep{PCW10, NS10} in which the evolution of the ejecta is approximately planar as long as its radius do not double. In the case of the NS-NS merger, since the initial length scale of the merger system is small $\sim10^6$ cm and the velocity of the merger ejecta is subrelativistic, the planar phase is irrelevant for the observations.

\subsection{Heating Mechanisms}

\subsubsection{Radioactivity}
\label{radio}

One of the two heating mechanisms we consider is nuclear heating by $r$-process elements. Since the beta decay products of $r$-process elements produced in NS binary mergers naturally heat ejecta, this mechanism is considered to power the emission of a macronova \citep[e.g., ][]{LP98}. The nuclear heating rate is calculated in several works \citep{Met+10, Rob+11, Kor+12, Ros+14, Wan+14}. The derived heating rates per unit mass $\dot{\epsilon}(t)$ are described by the following formula
\begin{eqnarray}\label{sec2-1:dotepsilon}
\dot{\epsilon}=\dot{\epsilon}_0\left(\frac{t}{1\ {\rm day}}\right)^{-\alpha}.
\end{eqnarray}
In this study, we use $\alpha=1.3$ and $\dot{\epsilon}_0=2\times10^{10}~{\rm erg~s}^{-1}{\rm g}^{-1}$ obtained by \citet{Wan+14}. The value of $\dot{\epsilon}$ has been obtained by simulations under some simplified assumptions with only limited parameter regions. Thus, we should note that the value of $\dot{\epsilon}_0$ has uncertainties. 

The injected internal energy by the nuclear decay is $\propto t^{1-\alpha}$ in the region $r < r_{\rm diff}$. On the other hand, the injected energy in this region is decreased by adiabatic cooling. The time-evolution of internal energy due to the adiabatic cooling is proportional to $t^{-1}$. Comparing the two temporal evolution, the index of the adiabatic cooling is smaller than that of the increase of internal energy due to the nuclear decay for $\alpha < 2$. Since we use $\alpha=1.3$, we neglect the injected internal energy in the region $r < r_{\rm diff}$. 

\subsubsection{Engine-driven shock}
\label{shock}

Unlike the $r$-process model, energy injection occurs only within the time $t_{\rm inj}$ in the engine model. We only consider adiabatic cooling as a cooling process of ejecta after $t_{\rm inj}$, and therefore, the temperature distribution at time $t$ is, 
\begin{eqnarray}\label{sec2-1:T}
T(t,v)=T_0\left(\frac{t}{t_{\rm inj}}\right)^{-1}\left(\frac{v}{v_{\rm min}}\right)^{-\xi},
\end{eqnarray}
where the index $\xi$ is a parameter for a snapshot distribution and $T_0$ is a normalization factor described later. The time dependence of $t^{-1}$ is the effect of adiabatic expansion. 

The normalized value $T_0$ is determined by using the relation of total injected internal energy $E_{\rm int0}$ as
\begin{eqnarray}\label{sec2-1:E_int}
E_{\rm int0}&=&4\pi\int_{v_{\min}t_{\rm inj}}^{v_{\max}t_{\rm inj}}aT^4(t_{\rm inj},v)r^2dr \nonumber \\
&=&\frac{4\pi}{3-4\xi}aT_0^4(v_{\min}t_{\rm inj})^3\left[\left(\frac{v_{\max}}{v_{\min}}\right)^{3-4\xi}-1\right],
\end{eqnarray}
where we use $dr=t_{\rm inj}dv$. For the temperature index $\xi>0.75$, the innermost region of ejecta has dominant internal energy. As will be shown in Section \ref{evolution}, since the luminosity and temperature always depend on the product of $E_{\rm int0}$ and $t_{\rm inj}$, we treat $E_{\rm int0}t_{\rm inj}$ as a parameter. Thus, the engine model has two parameters, $\xi$ and $E_{\rm int0}t_{\rm inj}$ instead of $\dot{\epsilon}_0$ and $\alpha$ in the $r$-process model.

Energy injection is not always a single event and the shock does not always get through the whole ejecta. It is considered that the activity of the central engine accompanies violent time variability. In this case, multiple shocks propagate into the ejecta. Some of the shock may not catch up with the outer edge of the ejecta. Current general relativistic simulations cannot calculate the evolution of ejecta for such a long time after merger ($t_{\rm inj}\sim 10^2$ s), so that the index $\xi$ of temperature distribution is highly uncertain. Therefore, we treat the temperature index $\xi$ as a parameter. 

Unlike the case of core-collapse supernova \citep{NS10}, it is difficult to determine the temperature distribution of heated ejecta by the activity of a central engine. In the case that the activity of a central engine injects the energy into the ejecta, the radiation-dominated shock (where the internal energy behind the shock is dominated by radiation) is formed in the ejecta. The ejecta are heated during the propagation of the shock. This situation is similar to the initial phase of core-collapse supernovae \citep[e.g., ][]{A80, P93}. In the cases of core-collapse supernovae, the kinetic energy of ejecta before the shock heating is much smaller than the injected internal energy. In such ejecta, the relation between velocity and mass density was obtained by \citet{S60} (in the non-relativistic case for the velocity of the ejecta). Using Sakurai's (1960) solution and the equipartition between the kinetic energy after the shock heating and the internal energy \citep{NS10}, the distribution of the temperature distribution is derived. However, in the case of compact binary mergers, the merger ejecta have a large velocity ($\sim0.01-0.1c$) before the shock heating \citep{Hot+13}. Then, injected internal energy is not always larger than the kinetic energy of ejecta so that it is not clear whether we can use the equipartition to estimate the distribution of internal energy or not.

The kinetic energy of the ejecta $E_{\rm kin}$ is described as 
\begin{eqnarray}\label{sec2-1:E_kin}
E_{\rm kin}&=&\frac{1}{2}\times4\pi\int_{v_{\min}}^{v_{\max}}\rho(t,v)v^4t^3dv \nonumber \\
&=&{\displaystyle \frac{1}{2}M_{\rm ej}v_{\min}^2\frac{(\beta-3)\left[\left(\frac{v_{\max}}{v_{\min}}\right)^{5-\beta}-1\right]}{(5-\beta)\left[1-\left(\frac{v_{\max}}{v_{\min}}\right)^{3-\beta}\right]}}.
\end{eqnarray}
Note that if the injected internal energy $E_{\rm int0}$ is larger than the kinetic energy of the ejecta, it is expected that some of the internal energy converts to the kinetic energy of the ejecta. As a result, the internal energy and the kinetic energy are equal as in the case of core-collapse supernovae. Then, the mass density distribution and the maximum velocity of the ejecta derived from simulations may be changed because the injection time may be long $\sim10^2$s compared to that calculated by simulations $\lesssim0.1$s \citep{Hot+13}. For simplicity, we only consider the case $E_{\rm int0}\le E_{\rm kin}$.

\section{Evolution of Luminosities and Temperatures}
\label{evolution}

\begin{table*}
\begin{center}
\begin{tabular}{clccc}
\multicolumn{5}{c}{TABLE 1 Model parameters.} \\ \hline
Symbol & & Fiducial model & Minimum mass model & Hot interior model \\ \hline
$M_{\rm ej}$ & Ejecta mass & $0.10M_{\odot}$ & $0.022M_{\odot}$ & $0.08M_{\odot}$ \\
$v_{\min}$ & Minimum velocity & $0.15c$ & $0.13c$ & $0.18c$ \\
$v_{\max}$ & Maximum velocity & $0.40c$ & $0.40c$ & $0.40c$ \\
$\beta$ & Index of the density profile & 3.5 & 3.5 & 3.5 \\
$\kappa$ & Opacity & 10 cm$^2$ g$^{-1}$ & 30 cm$^2$ g$^{-1}$ & 10 cm$^2$ g$^{-1}$ \\ 
$\dot{\epsilon}_0$ & Nuclear heating rate at 1 day & $2\times10^{10}$ erg s$^{-1}$ g$^{-1}$ & $\cdots$ & $2\times10^{10}$ erg s$^{-1}$ g$^{-1}$  \\ 
$\alpha$ & Index of nuclear heating rate & 1.3 & $\cdots$ & 1.3\\ 
$E_{\rm int0}$ & Internal energy at $t_{\rm inj}$ & $1.3\times10^{51}$ erg & $0.9\times10^{51}$ erg & $0.8\times10^{51}$ erg \\ 
$t_{\rm inj}$ & Injection time & $10^2$ s & $10^2$ s & $10^2$ s \\
$\xi$ & Index of the temperature profile & 1.6 & 1.1 & 2.7 \\ \hline
\multicolumn{5}{l}{}%
\end{tabular}
\label{tab:parameter}
\end{center}
\end{table*}

In this section, we present the evolution of the observed temperature and luminosity of a macronova using our model introduced in the previous section. In sections \ref{thin} -- \ref{transparent}, we focus on the parameter dependence of the evolution using some approximations. In section \ref{fiducial}, we calculate the temperature and luminosity using the fiducial model with parameters summarized in the first column of table 1. 

To calculate the luminosity and temperature, we assume that the emission is well described by the blackbody radiation \citep[e.g., ][]{BK13}. For simplicity, we assume that the observed temperature equals to the temperature at the diffusion radius $r_{\rm diff}$. We also assume that the temperature is not so different from the diffusion radius $r_{\rm diff}$ to $2r_{\rm diff}$ so that in the thick-diffusion phase ($r_{\rm out}>2r_{\rm diff}$), we only consider the emission from $r_{\rm diff}$ to $2r_{\rm diff}$ to calculate the observed luminosity for both the $r$-process and engine models. In some studies \citep[e.g., ][]{Met+14}, the observed temperature is approximated by the temperature at the radius of the photosphere $r_{\rm ph}$ where the optical depth is unity. Since the velocity of the ejecta is near the light speed, the optical depths at the diffusion radius $r_{\rm diff}$ and its twice $2r_{\rm diff}$ are $\tau\sim1-10^2$. Therefore, our assumed temperature approximately equals to the temperature at the photosphere. 

In section \ref{diffusion}, we introduced two phases, the thin- and thick-diffusion phases (figure \ref{figure:diffuse}), depending on the size of the region where photons make the diffusion in the ejecta $\Delta r$. We also introduce another phase $r_{\rm diff}\le r_{\rm in}$, the transparent phase, in which photons can diffuse out from the entire of the ejecta. Thus, we divide the evolution into these three phases for the values of the diffusion radius $r_{\rm diff}$ as described below. 

\subsection{Thin-diffusion phase}
\label{thin}

The size of the effectively thin region gets larger with time. At the early phase of a macronova, the diffusion radius $r_{\rm diff}$, which is the inner radius of the effectively thin region, is near the outer edge of the ejecta $r_{\rm out}$. In this early phase, we take the propagation distance $\Delta r$ of a photon as $\Delta r\sim r_{\rm out}-r_{\rm diff} (<r_{\rm diff})$. Since we assume that the density is a homologous function of the velocity $\rho\propto v^{-\beta}$, the density can be approximated as $\rho\sim\rho(v_{\max})$ in the region $r_{\rm diff}\gg\Delta r$. Using the escaping condition for the diffusing photons $t\sim t_{\rm diff}$, equation (\ref{sec2-1:t_diff}) and approximation on the optical depth $\tau\sim\Delta r\kappa\rho(t,v_{\max})$, the propagation distance $\Delta r$ can be estimated as
\begin{eqnarray}\label{sec2-2-1:Deltar}
\Delta r&\sim&\sqrt{\frac{ct}{\kappa\rho(t,v_{\max})}} \nonumber \\
&\propto&\kappa^{-1/2}M_{\rm ej}^{-1/2}v_{\min}^{\frac{3-\beta}{2}}v_{\max}^{\beta/2}t^2.
\end{eqnarray}
In the discussion of parameter dependence (sections \ref{thin} -- \ref{transparent}), we only consider the dominant term. For example, we neglect the second term in the right-hand side of equation (\ref{sec2-1:M_ej}) to derive the parameter equation (\ref{sec2-2-1:Deltar}) because the index of the mass density is $\beta>3$ in our model. In section \ref{fiducial}, we include the subdominant terms to calculate the light curves numerically.

First we consider the $r$-process model. The evolution of temperature $T_{\rm obs}$ is obtained by the internal energy density $\dot{\epsilon}t\rho$ at the radius $r=r_{\rm out}$. Using equations (\ref{sec2-1:rho}) and (\ref{sec2-1:M_ej}), the parameter dependence of the density is $\rho(t,v_{\max})\propto M_{\rm ej}v_{\min}^{\beta-3}v_{\max}^{-\beta}t^{-3}$. The observed temperature is 
\begin{eqnarray}\label{sec2-2-1:T_obs,nuc}
T_{\rm obs}&\sim&\left(\frac{\dot{\epsilon}t\rho(t,v_{\max})}{a}\right)^{1/4} \nonumber \\
&\propto&M_{\rm ej}^{1/4}v_{\min}^{\frac{\beta-3}{4}}v_{\max}^{-\beta/4}t^{-\frac{2+\alpha}{4}}.
\end{eqnarray}
For $\alpha=1.3$, the observed temperature evolves as $T_{\rm obs}\propto t^{-0.875}$. This is because in the thin-diffusion phase the ejecta is effectively a single expanding shell with $\rho\sim\rho(t,v_{\max})$ and the injected energy $\dot{\epsilon}t\propto t^{-0.3}$ is almost constant so that the observed temperature approximately follow adiabatic cooling $T\propto t^{-1}$. Note that in this phase the observed temperature does not depend on the opacity. The bolometric luminosity $L_{\rm bol}$ for the radioactivity is described as the product of the mass within the thickness $\Delta r$ in equation (\ref{sec2-2-1:Deltar}) and the nuclear heating rate $\dot{\epsilon}$ in equation (\ref{sec2-1:dotepsilon}) so that
\begin{eqnarray}\label{sec2-2-1:L_bol,nuc}
L_{\rm bol}&\sim& 4\pi r_{\rm out}^2\Delta r\rho(t,v_{\max})\dot{\epsilon} \nonumber \\
&\propto&\kappa^{-1/2}M_{\rm ej}^{1/2}v_{\min}^{\frac{\beta-3}{2}}v_{\max}^{\frac{4-\beta}{2}}t^{1-\alpha}.
\end{eqnarray}
For $\alpha=1.3$, the evolution of the bolometric luminosity is $L_{\rm bol}\propto t^{-0.3}$. 

Next we consider the engine model. We should take into account the freedom of the temperature index $\xi$ in the temperature distribution (equation \ref{sec2-1:T}). Since we only consider a dominant term in the right-hand side of equation (\ref{sec2-1:E_int}) (the first term for $\xi<0.75$ or the second term for $\xi > 0.75$) in this subsection, the parameter dependence of the temperature $T_0$ is described as 
\begin{eqnarray}\label{sec2-2-1:T_0}
T_0\propto E_{\rm int0}^{1/4}t_{\rm inj}^{-3/4}\times\left\{ \begin{array}{ll}
v_{\min}^{-\xi}v_{\max}^{\frac{4\xi-3}{4}} & (\xi < 0.75) \\
 & \\
v_{\min}^{-3/4} & (\xi > 0.75) \\
\end{array} \right. .
\end{eqnarray}
Substituting $v=v_{\max}$ into equation (\ref{sec2-1:T}), the observed temperature is described as
\begin{eqnarray}\label{sec2-2-1:T_obs,sh}
T_{\rm obs}&\sim&T_0\left(\frac{t}{t_{\rm inj}}\right)^{-1}\left(\frac{v_{\max}}{v_{\min}}\right)^{-\xi} \nonumber \\
&\propto&E_{\rm int0}^{1/4}t_{\rm inj}^{1/4}t^{-1}\times\left\{ \begin{array}{ll}
 v_{\max}^{-3/4} & (\xi < 0.75) \\
 & \\
 v_{\min}^{\frac{4\xi-3}{4}}v_{\max}^{-\xi} & (\xi > 0.75). \\
\end{array} \right.
\end{eqnarray}
Since the observed temperature $T_{\rm obs}$ approximately equals to the temperature at the outer edge of the ejecta, $T_{\rm obs}\sim T(t,v_{\max})$, the evolution of the observed temperature and the luminosity are also determined by the adiabatic cooling. The bolometric luminosity in the effectively thin region is equal to the total radiation created by thermal emission in this region \citep{RL79}. Using equation (\ref{sec2-2-1:T_obs,sh}), the bolometric luminosity is described as
\begin{eqnarray}\label{sec2-2-1:L_bol,sh}
L_{\rm bol}&\sim&4\pi r_{\rm out}^2\Delta r\frac{aT_{\rm obs}^4}{t} \nonumber \\
&\propto&\kappa^{-1/2}M_{\rm ej}^{-1/2}E_{\rm int0}t_{\rm inj}t^{-1} \nonumber \\
& &\times\left\{ \begin{array}{ll}
v_{\min}^{\frac{3-\beta}{2}}v_{\max}^{\frac{\beta-2}{2}} & (\xi < 0.75) \\
 & \\
v_{\min}^{\frac{-3-\beta+8\xi}{2}}v_{\max}^{\frac{4+\beta-8\xi}{2}} & (\xi > 0.75). \\
\end{array} \right.
\end{eqnarray}
The time evolution of bolometric luminosity for the engine model does not depend on the temperature index $\xi$ in the thin-diffusion phase. 

Comparing the engine model with the $r$-process model in the thin-diffusion phase, the bolometric luminosity and the observed temperature decrease faster in the engine model than those in the $r$-process model. These time-dependence do not depend on the indices of the density and temperature. 

Note that the light curve may depend on the detailed profile of the front of the ejecta in this thin-diffusion phase. The profile of the ejecta front is difficult to calculate by the numerical simulation due to its low density, and hence has large uncertain \citep{KIS14}. We discuss its dependence in section \ref{discuss}. 

\subsection{Thick-diffusion phase}
\label{thick}

We consider diffusion to evaluate the diffusion radius in the thick-diffusion phase. We take the propagation distance $\Delta r\sim r_{\rm diff}$ after the time when the difference between the radius of the outer edge of the ejecta $r_{\rm out}$ and the diffusion radius $r_{\rm diff}$ is larger than the diffusion radius, $r_{\rm out}-r_{\rm diff}>r_{\rm diff}$, since the optical depth of the outer part is negligible for the density profile in equation (\ref{sec2-1:rho}). In this thick-diffusion phase, mass density significantly deviates from $\rho(v_{\max})$. Substituting equations (\ref{sec2-1:t_diff}) and (\ref{sec2-1:tau2}) into $t=t_{\rm diff}$, the diffusion radius $r_{\rm diff}$ is calculated as
\begin{eqnarray}\label{sec2-2-2:r_diff}
r_{\rm diff}&\sim&\left[\frac{(\beta-3)\kappa M_{\rm ej}v_{\min}^{\beta-3}t^{\beta-4}}{4\pi(\beta-1)c}\right]^{\frac{1}{\beta-2}} \nonumber \\
&\propto&\kappa^{\frac{1}{\beta-2}}M_{\rm ej}^{\frac{1}{\beta-2}}v_{\min}^{\frac{\beta-3}{\beta-2}}t^{\frac{\beta-4}{\beta-2}},
\end{eqnarray}
where we use the relations $\Delta r\sim r_{\rm diff}$ and $v\sim r_{\rm diff}/t$. The latter is obtained from the assumption of the homologous expansion. Regarding the optical depth $\tau$, the second term is neglected in the right-hand side of equation (\ref{sec2-1:tau}) to focus only on the dominant term to study parameter dependence in sections \ref{thin} -- \ref{transparent}. For $\beta=3.5$, the diffusion radius decreases with time ($r_{\rm diff}\propto t^{-1/3}$). Then, emission from the region with relatively high mass density can be observed progressively in this phase ($\rho\propto t^{-3}(r_{\rm diff}/t)^{-\beta}\propto t^{(4\beta-9)/3}=t^{1.667}$). 

We introduce the transition time $t_{\times}$ between thin- and thick-diffusion phases, which satisfies the relation $r_{\rm diff}=0.5r_{\rm out}$. Substituting equation (\ref{sec2-2-2:r_diff}) and $r_{\rm out}=v_{\max}t$ into the relation $r_{\rm diff}=0.5r_{\rm out}$, we can obtain the transition time $t_{\times}$ as 
\begin{eqnarray}\label{sec2-2-2:t_times}
t_{\times}&\sim&\sqrt{\frac{2^{\beta-4}(\beta-3)\kappa M_{\rm ej}}{\pi(\beta-1)cv_{\max}}\left(\frac{v_{\max}}{v_{\min}}\right)^{3-\beta}} \nonumber \\
&\sim&4.1~\kappa_{10}^{1/2}M_{\rm ej,0.1}^{1/2}v_{\min,0.1}^{\frac{\beta-3}{2}}v_{\max,0.4}^{\frac{2-\beta}{2}}~{\rm day},
\end{eqnarray}
where $\kappa_{10}\equiv \kappa/10$ cm$^2$ g$^{-1}$, $M_{\rm ej,0.1}\equiv M_{\rm ej}/0.1M_{\odot}$, $v_{\min,0.1}\equiv v_{\min}/0.1c$ and $v_{\max,0.4}\equiv v_{\max}/0.4c$. As seen above, the transition time $t_{\times}$ is typically several days. This timescale is expected to allow the follow-up observations \citep{Aasi+14}. Thus, we should consider both phases to predict something useful for follow-up observations. If we fix $\kappa$ and $v_{\max}$ and use $\beta=3.5$, equation (\ref{sec2-2-2:t_times}) gives $t_{\times}\propto M_{\rm ej}^{1/2}v_{\min}^{1/4}$. If we increase the total mass of the ejecta $M_{\rm ej}$ and the velocity at the inner edge of the ejecta $v_{\min}$, the mass density of the ejecta $\rho$ and hence the optical depth are increased. As a result, the transition time $t_{\times}$ becomes large.  

First, we consider the $r$-process model. Here, we introduce the velocity $v_{\rm diff}=r_{\rm diff}/t$ based on the homologous relation. Using the velocity $v_{\rm diff}$ and equation (\ref{sec2-2-2:r_diff}) for the mass density (equation \ref{sec2-1:rho}), the evolution of temperature is
\begin{eqnarray}\label{sec2-2-2:T_obs,nuc}
T_{\rm obs}&\sim&\left(\frac{\dot{\epsilon}t\rho(t,v_{\rm diff})}{a}\right)^{1/4} \nonumber \\
&\propto&\kappa^{\frac{\beta}{4(2-\beta)}}M_{\rm ej}^{\frac{1}{2(2-\beta)}}v_{\min}^{\frac{\beta-3}{2(2-\beta)}}t^{\frac{1}{\beta-2}-\frac{\alpha}{4}}.
\end{eqnarray}
For $\alpha=1.3$ and $\beta=3.5$, the evolution is described by $T_{\rm obs}\propto t^{0.341}$ so that the observed temperature increases with time. The bolometric luminosity is described as the product of the mass between $r_{\rm diff}$ and $2r_{\rm diff}$ and the nuclear heating rate $\dot{\epsilon}$. Using equations (\ref{sec2-1:dotepsilon}) and (\ref{sec2-2-2:r_diff}), we obtain the bolometric luminosity as, 
\begin{eqnarray}\label{sec2-2-2:L_bol,nuc}
L_{\rm bol}&\sim&4\pi r_{\rm diff}^3\rho(t,v_{\rm diff})\dot{\epsilon} \nonumber \\
&\propto&\kappa^{\frac{3-\beta}{\beta-2}}M_{\rm ej}^{\frac{1}{\beta-2}}v_{\min}^{\frac{\beta-3}{\beta-2}}t^{\frac{2(\beta-3)}{\beta-2}-\alpha}.
\end{eqnarray}
For $\alpha=1.3$ and $\beta=3.5$, the evolution of the bolometric luminosity is $L_{\rm bol}\propto t^{-0.633}$. 

Next, we consider the engine model. Using equations (\ref{sec2-1:T}) and (\ref{sec2-2-2:r_diff}), the evolution of the observed temperature is described as
\begin{eqnarray}\label{sec2-2-2:T_obs,sh}
T_{\rm obs}&\sim&T_0\left(\frac{t}{t_{\rm inj}}\right)^{-1}\left(\frac{v_{\rm diff}}{v_{\min}}\right)^{-\xi} \nonumber \\
&\propto&\kappa^{-\frac{\xi}{\beta-2}}M_{\rm ej}^{-\frac{\xi}{\beta-2}}E_{\rm int0}^{1/4}t_{\rm inj}^{1/4}t^{\frac{-\beta+2\xi+2}{\beta-2}} \nonumber \\
& &\times\left\{ \begin{array}{ll}
 v_{\min}^{\frac{\xi(\beta-3)}{2-\beta}}v_{\max}^{\frac{4\xi-3}{4}} & (\xi < 0.75) \\
 & \\
 v_{\min}^{\frac{3\beta-6-4\xi}{4(2-\beta)}} & (\xi > 0.75). \\
\end{array} \right.
\end{eqnarray} 
For $\beta=3.5$, the value $\xi=0.75$ is the boundary whether the observed temperature increases with time ($\xi>0.75$) or not ($\xi<0.75$). The evolution of the luminosity equals to the total radiation created by thermal emission in the sphere with radius $r_{\rm diff}$. Using the relation $v(r_{\rm diff})\sim r_{\rm diff}/t$ and equations (\ref{sec2-2-2:r_diff}) and (\ref{sec2-2-2:T_obs,sh}), we obtain
\begin{eqnarray}\label{sec2-2-2:L_bol,sh}
L_{\rm bol}&\sim&4\pi r_{\rm diff}^3\frac{aT_{\rm obs}^4}{t} \nonumber \\
&\propto&\kappa^{\frac{3-4\xi}{\beta-2}}M_{\rm ej}^{\frac{3-4\xi}{\beta-2}}E_{\rm int0}t_{\rm inj}t^{\frac{2(\beta+1-4\xi)}{2-\beta}} \nonumber \\
& &\times\left\{ \begin{array}{ll}
v_{\min}^{\frac{(3-4\xi)(\beta-3)}{\beta-2}}v_{\max}^{4\xi-3} & (\xi < 0.75) \\
 & \\
v_{\min}^{\frac{3-4\xi}{2-\beta}} & (\xi > 0.75). \\
\end{array} \right.
\end{eqnarray}
If we take $\beta=3.5$ and $\xi=1.0$, the evolution of the bolometric luminosity is $L_{\rm bol}\propto t^{-0.666}$. This is almost the same dependence as in the $r$-process model. Note that even if the inner part of the ejecta has the larger internal energy ($\xi>0.75$), the bolometric luminosity does not always increase with time. Using the relation $E_{\rm int}(v,t)\propto t^{-1}$, from the adiabatic cooling, the evolution of bolometric luminosity for a given mass shell with $v$ is $L_{\rm bol}\sim E_{\rm int}(v)t^{-1}\propto t^{-2}$, where $E_{\rm int}(v,t)$ is the total internal energy for the mass shell with a given expanding velocity $v$. Since $E_{\rm int}(v_{\rm difff})\propto v^{3-4\xi}$ and $v_{\rm diff}= r_{\rm diff}/t\propto t^{\frac{2}{2-\beta}}$, the bolometric luminosity increases with time for the value of the temperature index $\xi>(\beta+1)/4=1.125$. 

\subsection{Transparent phase}
\label{transparent}

Once the diffusion radius reaches the inner edge of the ejecta $(r_{\rm diff}=r_{\rm in})$, all photons emitted from the ejecta can diffuse out within dynamical timescale. If energy is not injected into the ejecta in this transparent phase, the internal energy in the ejecta runs out immediately. The transition time from the thick-diffusion phase to the transparent phase $t_{\rm tr}$ is described as 
\begin{eqnarray}\label{sec2-2-3:t_tr}
t_{\rm tr}&\sim&\sqrt{\frac{(\beta-3)\kappa M_{\rm ej}}{4\pi(\beta-1)cv_{\min}}} \nonumber \\
&\sim&6.9~\kappa_{10}^{1/2}M_{\rm ej,0.1}^{1/2}v_{\min,0.1}^{-1/2}~{\rm day},
\end{eqnarray}
where we use the diffusion radius $r_{\rm diff}= r_{\rm in}$.

First we consider the $r$-process model. The observed temperature equals to the temperature at the inner edge of the ejecta, $T_{\rm obs}\sim [\dot{\epsilon}t\rho(v_{\min})/a]^{1/4}$. Using equation (\ref{sec2-1:rho}), we obtain
\begin{eqnarray}\label{sec2-2-3:T_obs,nuc}
T_{\rm obs}&\sim&\left(\frac{\dot{\epsilon}t\rho(t,v_{\min})}{a}\right)^{1/4} \nonumber \\
&\propto&M_{\rm ej}^{1/4}v_{\min}^{-3/4}t^{-\frac{2+\alpha}{4}}.
\end{eqnarray}
Since the energy is continuously injected due to the nuclear heating in the $r$-process model, the bolometric luminosity from the entire ejecta is described as $L_{\rm bol}\sim M_{\rm ej}\dot{\epsilon}$. However, the outer part of the ejecta emits photons with lower temperature and/or X-rays and $\gamma$-rays produced directly in radioactive decays. Although such emission contributes to the bolometric luminosity, we here focus only on the optical and infrared emissions. In the thick-diffusion phase, the observed emission comes from the region between $\sim r_{\rm diff}$ and $\sim2r_{\rm diff}$. In the transparent phase, we assume that the time evolution of the diffusion radius $r_{\rm diff}$ is the same as the thick-diffusion phase until $2r_{\rm diff}=r_{\rm in}$ and the observed luminosity comes from the region from $r_{\rm in}$ to $2r_{\rm diff}$ for simplicity. Then, the bolometric luminosity is described as
\begin{eqnarray}\label{sec2-2-3:L_bol,nuc}
L_{\rm bol}&\sim&4\pi r_{\rm in}^3\rho(t,v_{\min})\dot{\epsilon} \nonumber \\
&\propto&M_{\rm ej}t^{-\alpha}.
\end{eqnarray}
Although it appears that this time evolution directly reflects the nuclear decay rate, when we calculate the mass between $r_{\rm in}$ and $2r_{\rm diff}$ the evolution of the upper limit of the integration $2r_{\rm diff}$ makes the decrease of the luminosity faster than $\propto t^{-\alpha}$ (see a dashed line in the middle panel of figure \ref{figure:evolution}). In addition, the evolution of $r_{\rm diff}$ depends on the index $\beta$ (see equation \ref{sec2-2-2:r_diff}), so that the mass between $r_{\rm in}$ and $2r_{\rm diff}$ also depends on the index $\beta$. 

Next we consider the engine model. We assume that the internal energy is exhausted when the diffusion radius reaches $2r_{\rm diff}=r_{\rm in}$. For the observed temperature $T_{\rm obs}$, we assume the relation $T_{\rm obs}=T(t,v_{\min})$ and use equation (\ref{sec2-1:T}),
\begin{eqnarray}\label{sec2-2-3:T_obs,sh}
T_{\rm obs}&\sim&T_0\left(\frac{t}{t_{\rm inj}}\right)^{-1} \nonumber \\
&\propto&E_{\rm int0}^{1/4}t_{\rm inj}^{1/4}t^{-1} \nonumber \\
& &\times\left\{ \begin{array}{ll}
 v_{\min}^{-\xi}v_{\max}^{\frac{4\xi-3}{4}} & (\xi < 0.75) \\
 & \\
 v_{\min}^{-3/4} & (\xi > 0.75). \\
\end{array} \right.
\end{eqnarray}
The bolometric luminosity is described as
\begin{eqnarray}\label{sec2-2-3:L_bol,sh}
L_{\rm bol}&\sim&4\pi \int_{r_{\rm in}}^{2r_{\rm diff}}\frac{aT_{\rm obs}^4}{t} \nonumber \\
&\propto&E_{\rm int0}t_{\rm inj}t^{-2} \nonumber \\
& &\times\left\{ \begin{array}{ll}
\kappa^{\frac{3-4\xi}{\beta-2}}M_{\rm ej}^{\frac{3-4\xi}{\beta-2}}v_{\min}^{\frac{(\beta-3)(3-4\xi)}{\beta-2}} & \\
~~\times v_{\max}^{4\xi-3}t^{\frac{2(3-4\xi)}{2-\beta}} & (\xi < 0.75) \\
 & \\
1 & (\xi > 0.75). \\
\end{array} \right.
\end{eqnarray}
Since the internal energy at the innermost region almost equals to the total internal energy $E_{\rm int}(v_{\min})\sim E_{\rm int0}(t/t_{\rm inj})^{-1}$ and determines the bolometric luminosity $L_{\rm bol}\sim E_{\rm int}(v_{\min})/t$ for the temperature index $\xi>0.75$, the bolometric luminosity does not depend on the mass $M_{\rm ej}$ and velocities $v_{\max}$ and $v_{\min}$. This luminosity always corresponds to the maximum luminosity for $\xi > 0.75$, so that we can impose the lower limit on the parameter $E_{\rm int0}t_{\rm inj}$. 

\subsection{Fiducial Model}
\label{fiducial}

　 \begin{figure}
  \begin{center}
   \includegraphics[width=70mm]{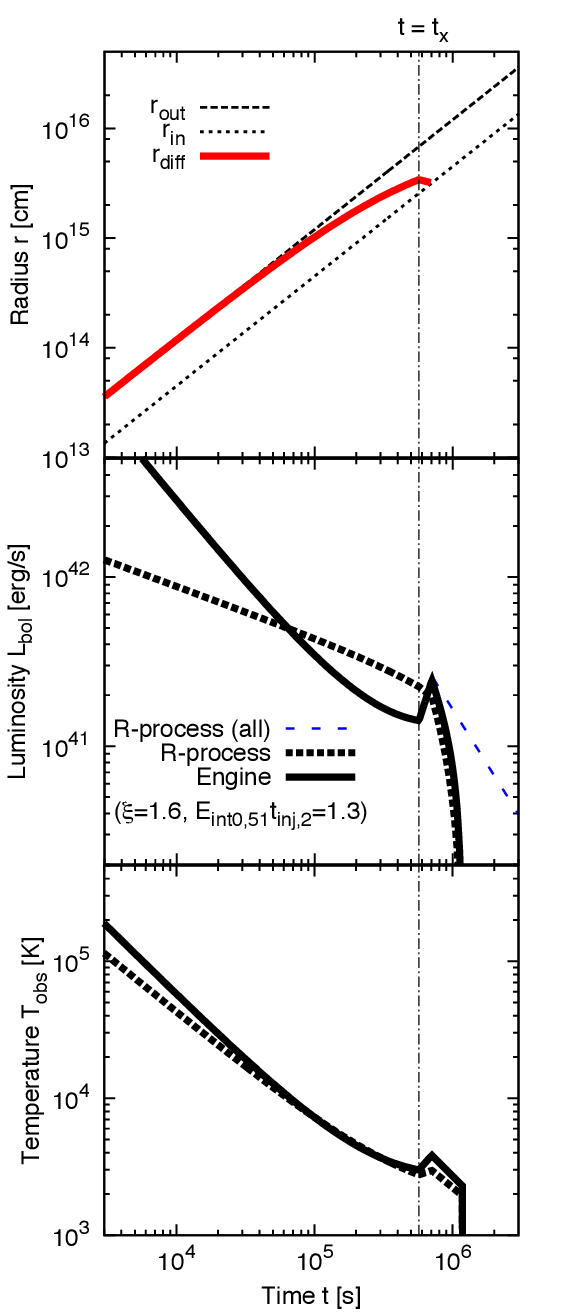}
   \caption{ Temporal evolution of the diffusion radius (top), bolometric luminosities (middle) and observed temperatures (bottom) in the fiducial model (first column of table 1). Thick dashed and solid lines show the evolution for the $r$-process model and the engine model, respectively. For comparison, we also plot the bolometric luminosity from the whole ejecta for the $r$-process model after the transparent phase ($t>t_{\rm tr}$ in equation \ref{sec2-2-3:t_tr}) as a blue long-dashed line in the middle panel.}
   \label{figure:evolution}
  \end{center}
 \end{figure}

 \begin{figure*}
  \begin{center}
   \includegraphics[width=120mm]{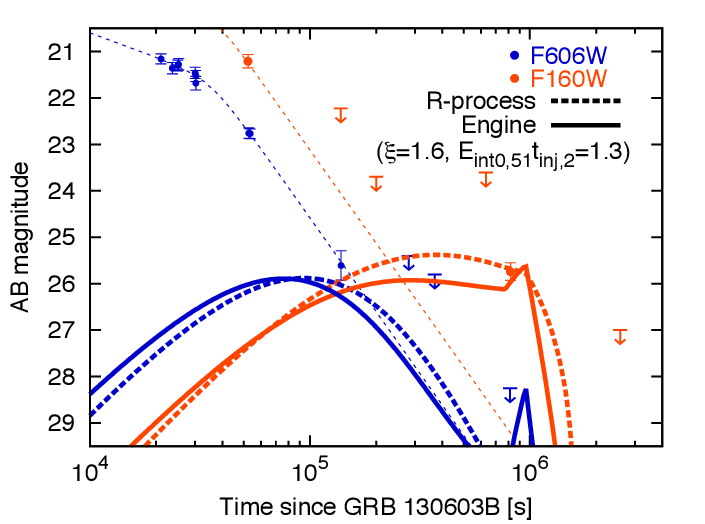}
   \caption{Theoretical light curves calculated under the fiducial parameter set (Table 1) at a near-infrared band (F160W, {\it red}) and optical band (F606W, {\it blue}). Two models (the $r$-process model, {\it solid}; the engine model, {\it dashed}) are considered. The observational results of GRB 130603B \citep[z = 0.356; ][]{Tan+13, BFC13, Cuc13, de14} are also plotted. The thin dotted lines are light curves calculated from a GRB afterglow model \citep{Tan+13}. Both models can reproduce the observational data well. }
   \label{figure:observation}
  \end{center}
 \end{figure*}

We show the temporal evolution of the diffusion radius $r_{\rm diff}$, the bolometric luminosity $L_{\rm bol}$ and the observed temperature $T_{\rm obs}$ in figure \ref{figure:evolution} under the fiducial parameter set. The parameters are summarized in the first column of table 1. Here, we do not use approximations $\rho(v)\sim\rho(v_{\rm max})$ and $T(v)\sim T(v_{\max})$ at the thin-diffusion phase as in section \ref{thin}. Instead, the diffusion radius $r_{\rm diff}$ is calculated from equations (\ref{sec2-1:t_diff}) -- (\ref{sec2-1:tau}) without approximations. Using the obtained diffusion radius $r_{\rm diff}$ and the relation $v_{\rm diff}=r_{\rm diff}/t$, we calculate the observed temperatures in the thin- and thick-diffusion phases, $T_{\rm obs}\sim[\dot{\epsilon}t\rho(t,v_{\rm diff})/a]^{1/4}$ (equation \ref{sec2-2-2:T_obs,nuc}), and $T_{\rm obs}\sim T_0(t/t_{\rm inj})^{-1}(v_{\rm diff}/v_{\min})^{-\xi}$ (equation \ref{sec2-2-2:T_obs,sh}) for the $r$-process and the engine models, respectively. In the transparent phase, the temperature in equations (\ref{sec2-2-3:T_obs,nuc}) and (\ref{sec2-2-3:T_obs,sh}) are evaluated with $v=v_{\min}$. Equations on observed temperature and bolometric luminosity for both models are summarized in appendix. The set of parameters we choose here explains the observed optical and infrared light curves of GRB 130603B (see next section). The vertical dash-dotted lines in figure \ref{figure:evolution} show the time $t=t_{\times}$ (equation \ref{sec2-2-2:t_times}). The diffusion radius is plotted only up to the transition time $t=t_{\rm tr}$ (equation \ref{sec2-2-3:t_tr}). 

In the thick-diffusion phase, the diffusion radius ($r_{\rm diff}\propto t^{\frac{\beta-4}{\beta-2}}=t^{-1/3}$ for $\beta=3.5$) moves inward in the ejecta ($r\propto t$). Since the observed luminosity and temperature are determined at the diffusion radius $r_{\rm diff}$, the time evolution of luminosity and temperature strongly depends on the indices of the profile, $\beta$ and $\xi$. For the $r$-process model, the bolometric luminosity decreases with time ($L_{\rm bol}\propto t^{\frac{2(\beta-3)}{\beta-2}-\alpha}=t^{-0.633}$) in the thick-diffusion phase (see equation \ref{sec2-2-2:L_bol,nuc}), which is more rapid than that in the thin-diffusion phase ($L_{\rm bol}\propto t^{1-\alpha}=t^{-0.3}$, see equation \ref{sec2-2-1:L_bol,nuc}). Since the index of the mass density $\beta=3.5$ is close to 3, in which the mass of each shell with a certain size $\delta r$ is the same value in logarithmic scale, the mass between the diffusion radius $r_{\rm diff}$ and its doubled value $2r_{\rm diff}$ does not significantly change with time. The luminosity is mainly determined by that mass, so that the evolution of the luminosity is slow compared with the evolution of nuclear heating rate ($\propto t^{-\alpha}$) in the thick-diffusion phase. On the other hand, bolometric luminosity and observed temperature increase with time in the engine model with the parameter set of the fiducial model. These mainly reflect the profile of the temperature distribution ($\xi=1.6$). In fact, using equation (\ref{sec2-2-2:L_bol,sh}), the index of the time $t$ for the bolometric luminosity is $-2(\beta+1-4\xi)/(\beta-2)\sim2.53$ for the engine model. 

After the transition time $t\ge t_{\rm tr}$, the luminosity and temperature are almost determined by the quantities at the inner edge of the ejecta. Then, the evolution of the luminosity and temperature does not significantly depend on the indices of profile $\beta$ and $\xi$ as in the case of the thin-diffusion phase (except for the case $\xi<0.75$ of the engine model, equation \ref{sec2-2-3:L_bol,sh}). Since our used profile of mass density has an artificially steep cut-off at the inner edge of the ejecta (figure \ref{figure:diffuse}), bolometric luminosity in both models rapidly declines after the time $t\ge t_{\rm tr}$. In the bottom panel of figure \ref{figure:evolution}, the observed temperature in both models has a steep cutoff at $2r_{\rm diff}=r_{\rm in}$. For comparison, we also consider the time evolution of bolometric luminosity from the whole ejecta $L_{\rm bol}=M_{\rm ej}t^{-\alpha}$ in the $r$-process model. Time evolution is shown in the middle panel of figure \ref{figure:evolution} as a blue long-dashed line. This luminosity evolution ($L_{\rm bol}\propto t^{-\alpha}=t^{-1.3}$) is significantly slower than that of the engine model in the transparent phase. In section \ref{implication}, we discuss the implication for discriminating the $r$-process model and the engine model using these temporal behaviors.

\section{DISCUSSION}
\label{discuss}

 \begin{figure}
  \begin{center}
   \includegraphics[width=100mm]{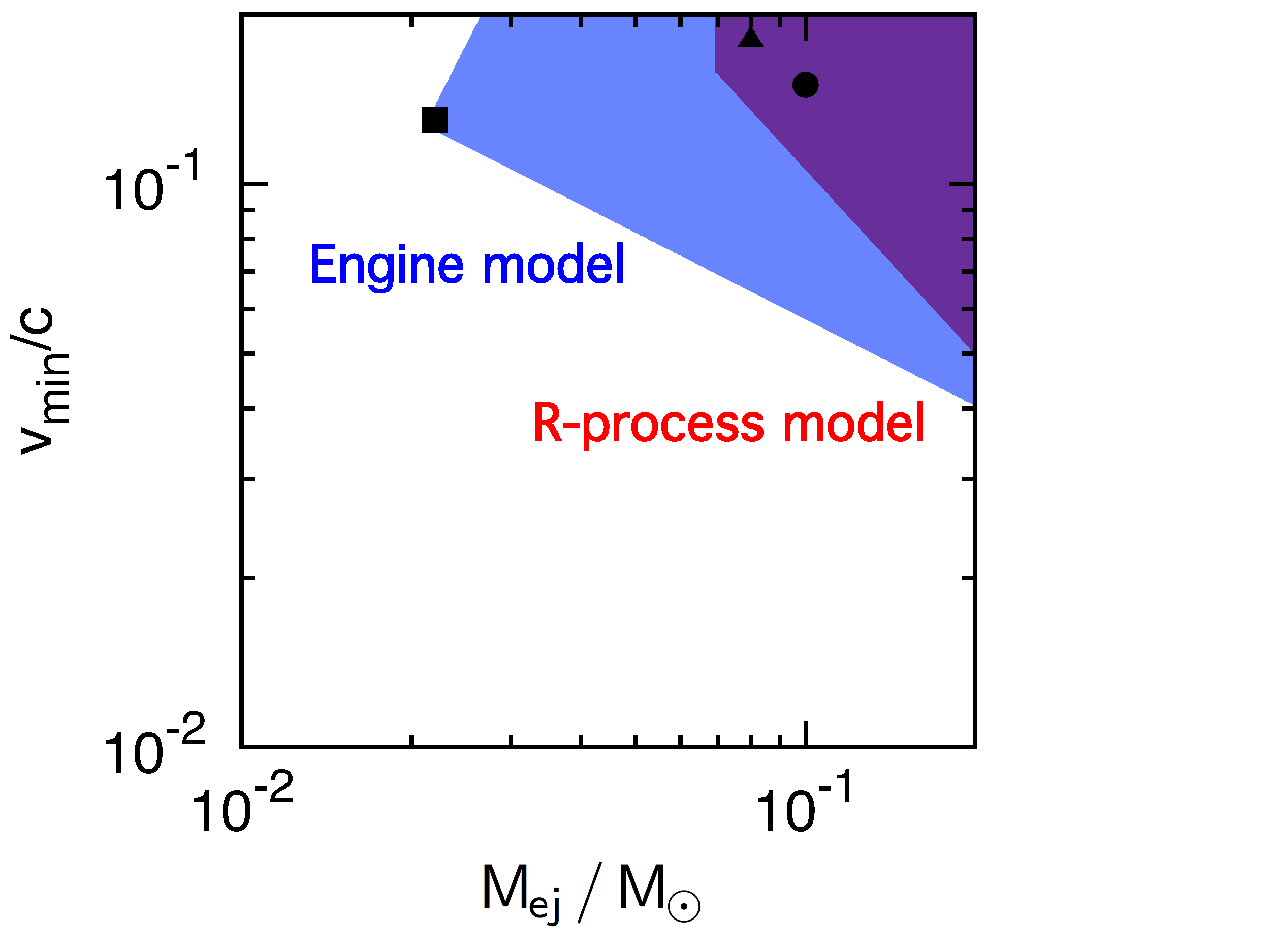}
   \caption{Range of parameter space in order to explain the observations of the macronova, GRB 130603B ({\it blue} for the engine model and {\it red} for the $r$-process model). Since two regions are overlapped, the color looks like purple for the $r$-process model. These regions are only a schematic view. We fix $v_{\max}=0.4c$, $\beta=3.5$, $\dot{\epsilon}_0=2\times10^{10}$ erg s$^{-1}$ g$^{-1}$ and $\alpha=1.3$. For the opacity $\kappa$, we use the range $\kappa=3-30$ cm$^2$ g$^{-1}$. For the engine model, we treat $E_{\rm int0}t_{\rm inj}$ and $\xi$ as free parameters to fit the light curve. The circle, square and triangle denote the case of the fiducial model, minimum mass model and hot interior model, respectively (see table 1).}
   \label{figure:parameter}
  \end{center}
 \end{figure}

 \begin{figure*}
  \begin{center}
   \includegraphics[width=170mm]{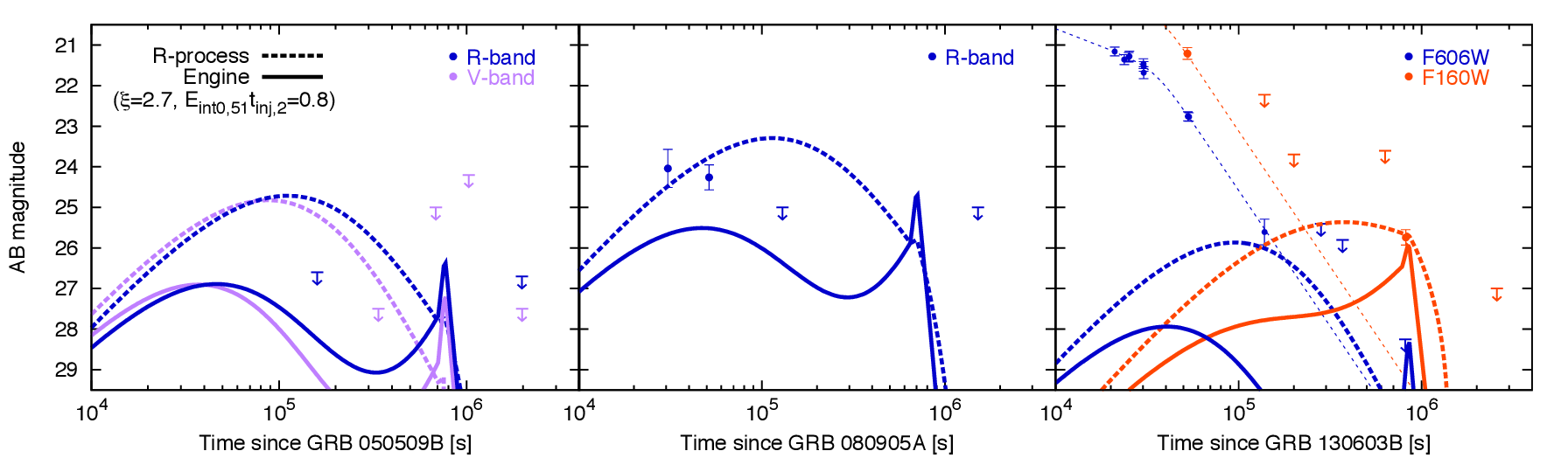}
   \caption{Theoretical light curves calculated under the hot interior model (table 1) at V ({\it purple}), R, F606W ({\it blue}) and F160W bands ({\it red}). Two models (the $r$-process model, {\it solid}; the engine model, {\it dashed}) are considered. The observational results of GRB 050509B \citep[left panel, $z=0.122$, ][]{Hjo+05}, GRB 080905A \citep[middle panel, $z=0.225$, ][]{Row+10} and GRB 130603B \citep[right panel, $z=0.356$, ][]{Tan+13, BFC13, Cuc13, de14} are also plotted. The engine model can reproduce all three observational data well.}
   \label{figure:observation2}
  \end{center}
 \end{figure*}

\subsection{Comparison with GRB 130603B}

We compare the results with the optical and infrared observations of short GRB 130603B in figure \ref{figure:observation}. The fiducial parameter set in table 1 is adopted. The $r$-process model and the engine model result in similar light curves at the optical and infrared bands. Both of them satisfy the observational data of GRB 130603B. Note that the detection point at F606W band at $\sim10^5$ s is consistent with the afterglow of GRB 130603B modeled as a smoothly broken power law \citep[blue dashed line, ][]{Tan+13}. We regard this detected value as an upper limit for the luminosity of emission from the ejecta. The detection point at F160W band at $\sim10^6$ s exceeds the extrapolation of the afterglow emission \citep[red dashed line, ][]{Tan+13}, so that we regard this detected emission as a thermal radiation from the ejecta. 

The range of the model parameters $v_{\min}$ and $M_{\rm ej}$ to satisfy the constraints obtained from the observation of GRB 130603B is shown in figure \ref{figure:parameter} as colored areas (red area for the $r$-process model and blue area for the engine model). Note that the red area has a completely overlap with the blue area. We fix the other model parameters $v_{\max}=0.4c$, $\beta=3.5$, $\dot{\epsilon}_0=2\times10^{10}$ erg s$^{-1}$ g$^{-1}$ and $\alpha=1.3$ as in the fiducial model. We take into account the uncertain range of the opacity, $\kappa=3-30$ cm$^2$ g$^{-1}$ to constrain the parameters, $v_{\min}$ and $M_{\rm ej}$. In the engine model, $\xi$ and $E_{\rm int0}t_{\rm inj}$ are additionally treated as free parameters to derive the allowed area in figure \ref{figure:parameter}.

\subsubsection{Limits on ejecta mass}

In the $r$-process model, the luminosity becomes smaller for smaller ejecta mass $M_{\rm ej}$. The small ejected mass $M_{\rm ej}\lesssim0.07M_{\odot}$ cannot reproduce the infrared excess of GRB 130603B (figure \ref{figure:observation}). The required ejecta mass is relatively large compared to the mass indicated by recent numerical simulations for a merger of binary NSs \citep[e.g., ][]{Hot+13, Ros+14, Jus+14}. Note that in \citet{BFC13}, $0.03 - 0.08 M_{\odot}$ is required to explain the observed infrared excess, which is a factor $\sim$2 smaller than our results. Their theoretical light curves are based on the study of \citet{BK13}. In \citet{BK13}, a broken power-law mass density profile with the index $-1$ for the inner layer and $-10$ for the outer layer of ejecta is adopted, in which the mass of the ejecta is efficiently concentrated at the transition point of the density index. Therefore, the luminosity of ejecta is evaluated as the heating rate multiplied by the total ejecta mass at the moment when the diffusion radius reaches the transition point. On the other hand, the index of our mass density profile of the ejecta is $\beta = 3.5$, which is indicated by general relativistic simulations by \citet{Hot+13}. This profile is quite different from the profile adopted in \citet{BK13}; the index is close to 3, in which the mass in each logarithmic radius is constant. Then, at the time $t = t_{\rm tr}$ (equation \ref{sec2-2-3:t_tr}), the mass contributing to the luminosity is about $\sim$60\% of the total ejecta mass in the case $v_{\min}=0.1c$. This profile predicts luminosity dimmer than that other studies. In fact, \citet{Hot+13b} tried to explain the observed infrared excess using the mass profile which is almost the same with ours. In the case of a binary NS merger with ejecta mass $\sim$0.02 $M_{\odot}$, even if they use a larger nuclear heating rate (larger by a factor of 2), their predicted luminosity is slightly smaller than the observed infrared excess (in the left panel of their figure 3). This result is consistent with our model, i.e., our model requires larger mass than most of previous studies. 

In the engine model, the injected internal energy which determines the luminosity does not depend on the ejecta mass (except for the limit in equation \ref{sec2-1:E_int}). However, the luminosity declines rapidly after the transition time $t\gtrsim t_{\rm tr}$ which depends on the ejecta mass as in equation (\ref{sec2-2-3:t_tr}). The condition $t_{\rm tr}\gtrsim10^6$ s in the observer frame is required to reproduce the excess observed from GRB 130603B in the near-infrared band. This condition gives the lower limit for the ejecta mass in the engine model, $M_{\rm ej}\gtrsim0.02M_{\odot}$ with the opacity $\kappa\sim30$ cm$^2$ g$^{-1}$. 

Note that the observed upper limit on the infrared luminosity at $\sim3\times10^6$ s in the observer frame (figure \ref{figure:observation}), which corresponds to $t_{\rm tr}\lesssim3\times10^6$ s, gives the upper limit on the ejecta mass for both models. However, this limit is not important for the range $M_{\rm ej}<0.2M_{\odot}$ in the range of the opacity $\kappa=3-30$ cm$^2$ g$^{-1}$.

\subsubsection{Limits on the minimum velocity}

The smaller minimum velocity $v_{\min}$ gives the smaller bolometric luminosity at certain time in the $r$-process model (see equations \ref{sec2-2-1:L_bol,nuc} and \ref{sec2-2-2:L_bol,nuc}). The small minimum velocity enlarges the size of ejecta (when we fix the maximum velocity $v_{\max}$). Then, the diffusion time $t_{\rm diff}$ of photons emitted from the inner region of the ejecta becomes long for the small velocity $v_{\min}$ (equation \ref{sec2-2-3:t_tr}). The mass between $r_{\rm diff}$ and $2r_{\rm diff}$ (or $r_{\rm out}$) increases toward inner region of the ejecta (as long as $\beta>3$) so that the mass is reduced for the small minimum velocity $v_{\rm min}$ at certain time. In fact, the dependence of the mass on the minimum velocity is $4\pi r_{\rm diff}^3\rho(t,v_{\rm diff})\propto v_{\min}^{\frac{\beta-3}{\beta-2}}=v_{\min}^{1/3}$. As a result, smaller minimum velocity gives smaller luminosity to reproduce the observed infrared excess of GRB 130603B. Moreover, smaller minimum velocity gives larger temperature $T_{\rm obs}$ at certain time (equations \ref{sec2-2-2:T_obs,nuc} and \ref{sec2-2-3:T_obs,nuc}) because mass density at a shell with small velocity is large. The difference between the detected luminosity at F160W band and the upper limit on the luminosity at F606W band at $\sim10^6$ s in the observer frame gives the upper limit on the observed temperature ($T_{\rm obs}\lesssim4\times10^3$ K). To satisfy the observed upper limit on the temperature from GRB 130603B, a lower limit of $v_{\min}\gtrsim0.1c$ is obtained for $M_{\rm ej}\sim0.1M_{\odot}$. 

The smaller minimum velocity $v_{\min}$ gives higher temperature $T_{\rm obs}$ in the engine model (equations \ref{sec2-2-2:T_obs,sh} and \ref{sec2-2-3:T_obs,sh}). The observational limit for the temperature at $\sim10^6$ s in the observer frame indicates that the range of the minimum velocity $v_{\rm min}$ is limited in the engine model ($v_{\min}\gtrsim0.06c$ for $M_{\rm ej}\sim0.1M_{\odot}$). 

\subsubsection{Dependence on opacity}

We discuss the dependence on the value of $\kappa$. As mentioned in section 2.2, we use the temperature-independent opacity $\kappa$ with the grey approximation. In general, the $r$-process line opacity depends on frequency and changes with temperature and ionization state of the ejecta \citep{KBB13, TH13}. The indicated grey opacity is $\kappa=3-30$ cm$^2$ g$^{-1}$. 

In the case of the $r$-process model, the luminosity significantly depends on opacity $\kappa$. The larger opacity causes larger diffusion time $t_{\rm diff}$, so that larger time is required to observe the inner region of the ejecta for given ejecta mass $M_{\rm ej}$ and minimum velocity $v_{\min}$. In fact, two transition times $t_{\times}$ and $t_{\rm tr}$ are proportional to $\kappa^{1/2}$ (equations \ref{sec2-2-2:t_times} and \ref{sec2-2-3:t_tr}). Then, the mass around the diffusion radius $r_{\rm diff}$ is small at certain time, so that the luminosity is reduced. As a result, in order to explain the infrared excess observed in GRB 130603B, larger mass $M_{\rm ej}$ is required for the larger value of opacity $\kappa$. For the opacity $\kappa>30$ cm$^2$ g$^{-1}$, total ejecta mass $M_{\rm ej}\gtrsim0.2M_{\odot}$ is required to reproduce the observed excess, which is much larger than the simulation results of mergers of binary NSs \citep[e.g., ][]{Hot+13}. On the other hand, the transition time $t_{\rm tr}$ is smaller for the smaller value of the opacity. Then, the luminosity significantly increases at $\sim10^5$ s. For the opacity $\kappa\lesssim3$ cm$^2$ g$^{-1}$, there is no parameter set which gives smaller luminosity than the detection at F606W band ($\sim10^5$ s in the observer frame) and the luminosity comparable to the observed excess at F160W band simultaneously in the $r$-process model. 

In the case of the engine model, a larger value of opacity $\kappa$ reduces the lower limit for the mass $M_{\rm ej}$ to explain the observed excess. For certain temperature and luminosity, the opacity $\kappa$ and the ejecta mass $M_{\rm ej}$ always degenerate in the form $\kappa M_{\rm ej}$ (see equations \ref{sec2-2-1:T_obs,sh}, \ref{sec2-2-1:L_bol,sh}, \ref{sec2-2-2:T_obs,sh}, \ref{sec2-2-2:L_bol,sh}, \ref{sec2-2-3:T_obs,sh} and \ref{sec2-2-3:L_bol,sh}). This dependence comes from the optical depth (equation \ref{sec2-1:tau}) because the internal energy in the ejecta does not depend on the opacity and the ejecta mass, contrary to the $r$-process model. We present a parameter set to give the minimum ejecta mass $M_{\rm ej}$ in table 1 as the minimum mass parameter set. We also plot the value of $M_{\rm ej}$ and $v_{\min}$ of this model in figure \ref{figure:parameter} as a square. This ejecta mass is naturally realized in general relativistic simulations \citep[e.g., ][]{Hot+13}. Although the larger value of the opacity $\kappa$ reduces the lower limit for the ejecta mass $M_{\rm ej}$, the kinetic energy is $E_{\rm kin}\sim1.1\times10^{51}$ erg (equation \ref{sec2-1:E_kin}) which is close to the initial injected energy $E_{\rm int0}=0.9\times10^{51}$ erg for the minimum mass parameter set. The lower ejecta mass $M_{\rm ej}$ reduces the kinetic energy of the ejecta, $E_{\rm kin}(\propto M_{\rm ej}v_{\max}^{5-\beta}v_{\min}^{\beta-3})$ in equation (\ref{sec2-1:E_kin}), so that the required energy $E_{\rm int0}$ may exceed the kinetic energy of the ejecta for the larger opacity. 
 For the small value of the opacity, larger mass and smaller minimum velocity is required to satisfy the condition $t_{\rm tr}\gtrsim10^6$ s ($t_{\rm tr}\propto\kappa^{1/2}M_{\rm ej}^{1/2}v_{\min}^{-1/2}$ in equation \ref{sec2-2-3:t_tr}). For the opacity $\kappa=3$ cm$^2$ g$^{-1}$, the condition corresponds to $(M_{\rm ej}/0.2M_{\odot})(v_{\min}/0.1c)^{-1}\gtrsim1$. The observational constraint for the temperature also requires a large value of the minimum velocity $v_{\min}$. Then, there is no solution to explain the observed excess within the parameter range shown in figure \ref{figure:parameter} for the opacity $\kappa\le3$ cm$^2$ g$^{-1}$. Therefore, for the small opacity $\kappa\le 3$ cm$^2$ g$^{-1}$ the engine model cannot explain the observed excess.

\subsubsection{Dependence on engine parameters}

Since the engine model has additional free parameters, $\xi$ and $E_{\rm int0}t_{\rm inj}$, the allowed region of the parameters is larger than that of the $r$-process model. We can impose the lower limit on the parameter $E_{\rm int0}t_{\rm inj}$ by regarding the infrared luminosity $\sim10^{41}$ erg s$^{-1}$ at $t\sim 7$ day as bolometric luminosity in the source rest frame with equation (\ref{sec2-2-3:L_bol,sh}). The derived limit is $(E_{\rm int0}/10^{51}{\rm erg})(t_{\rm inj}/10^2{\rm s})\gtrsim0.4$. To satisfy the optical upper limit at $\sim10^5$ s and the detected luminosity at $\sim10^6$ s in the observer frame (figure \ref{figure:observation}), we find the lower limit on the index of the temperature profile $\xi\gtrsim1.0$. For a smaller value of the index $\xi$, emission from the ejecta with relatively high temperature can be observed at time $\sim10^5$ s, so that luminosity at F606W band is larger than the observed upper limit of GRB 130603B. In addition, the smaller value of $\xi$ decreases relative internal energy in the inner edge of the ejecta. To reproduce the luminosity at time $\sim 10^6$ s in the observer frame when observed emission comes from the inner ejecta, the smaller $\xi$ requires the larger initial internal energy $E_{\rm int0}$ which exceeds the kinetic energy of the ejecta $E_{\rm kin}$ in some cases. 

\subsection{Comparison with Other GRBs with Deep Optical Observations}
\label{other}

Several deep optical observations of short GRBs give stringent upper limits on the luminosity of macronovae \citep{Kann+11}. We compare the results with two deep optical observations of short GRBs, GRB 050509B and GRB 080905A. For the fiducial parameter set, the luminosity exceeds the observational upper limits on these two observations. In the engine model, we can reduce the luminosity in the early phase $\lesssim10^5$ s without reducing the luminosity in the late phase $\sim10^6$ s by utilizing the steep temperature profile (large $\xi$). 

Here, we introduce the hot interior parameter set with larger value of index $\xi$ than that of the fiducial parameter set. Since emission from the inner part of the ejecta is observed at the later time, the luminosity at the early phase decreases and avoids the observational limits if most of the internal energy is injected to the inner part of the ejecta. We show the light curve of the hot interior parameter set in figure \ref{figure:observation2}. We choose the parameters as $M_{\rm ej}=0.08M_{\odot}, v_{\min}=0.18c, t_{\rm inj}=10^2{\rm s}, E_{\rm int0}=0.8\times10^{51}{\rm erg}$ and $\xi=2.7$ (the right column of table 1). From figure \ref{figure:observation2}, the light curves are consistent with all three observations using the same model parameters. A possible scenario for the hot interior parameter set is that the shock produced by the activity of the central engine may not be able to catch up with the outer part of the ejecta because the velocity of the ejecta is close to the light speed ($v_{\max}=0.4c$). Then, only the inner part of the ejecta will be heated. For comparison, we also show the light curves in the $r$-process model with the parameter set of the hot interior in figure \ref{figure:observation2} as dashed lines. The luminosity of the $r$-process model exceeds the observed upper limits in two observations, GRB 050509B and GRB 080905A (left and middle panels of figure \ref{figure:observation2}) if we choose the parameter set of the hot interior model. We are not able to find any parameter set in the $r$-process model, which simultaneously satisfies the observed limits of the three observations. Note that we do not argue that the $r$-process model is excluded from these results because we need to take into account the variations of the model parameters for each event.

Note that the extended emission was not detected in three short GRBs. However, it is not unreasonable to miss the extended emission of these bursts. One possibility is a selection effect. Observationally, the fraction of short GRBs with extended emission is significantly larger at softer energy bands: $\sim$25\% in the Swift BAT samples \citep[$>$15 keV; ][]{NGS10} and $\sim$7\% in the BATSE samples \citep[$>$20 keV; ][]{BKG13}. This suggests that observations with a low energy threshold may dramatically increase short GRBs with extended emission \citep{Nak+13}. This will be further tested by future soft X-ray survey facilities such as Wide-Field MAXI (0.7-10 keV) \citep{Kaw+14}. The three referred short GRBs were detected by Swift BAT and therefore Swift BAT could not detect extended emission by chance. Alternatively, the outflow following the main short GRB jet could not breakout the ejecta. \citet{Nag+14} and \citet{Mur+14} investigated the propagation of jets in merger ejecta. They found the cases that relativistic jets can penetrate merger ejecta and produce the prompt emission of short GRBs, but in the late energy injection cases, outflow fails to breakout the ejecta. Therefore, some extended emission may not be observed, although the central engine works actively.

\subsection{Outer Region of Mass Density Profile}
\label{outer}

\begin{figure}
  \begin{center}
   \includegraphics[width=80mm]{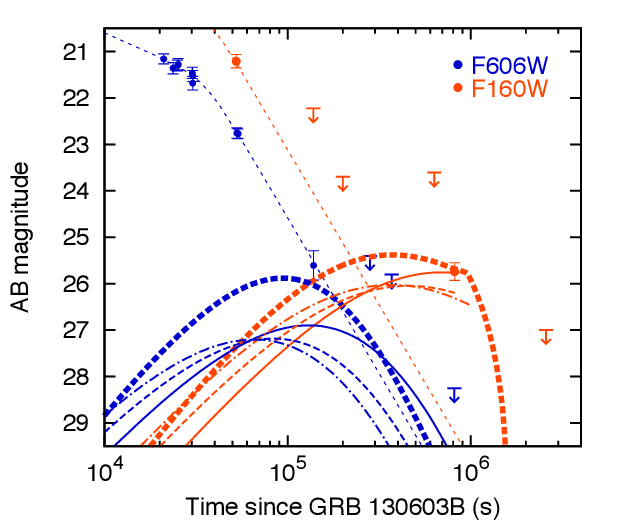}
   \caption{Dependence of theoretical light curves for the $r$-process model on the shape of the front of the ejecta. We plot the case of GRB 130603B. Thick-dashed lines denote the fiducial model. Thin-solid, thin-dashed, and thin-dot-dashed lines correspond to the exponential profiles of mass density (equation \ref{sec3:rho}) with $v_{\max}' = 0.4c, 0.5c$ and $0.6c$, respectively. We also plot the observational data from \citet{Tan+13, BFC13, Cuc13, de14}.}
   \label{figure:exp_comp}
  \end{center}
 \end{figure}

In the thin-diffusion phase, the light curve strongly depends on the density profile of the ejecta surface. The density profile is determined by the complex merger dynamics \citep{Hot+13}, so that the density profile of the ejecta cannot be analytically derived as mentioned in section 2.3. Since the outer part of the density profile is difficult to calculate precisely, little attention has paid on the mass profile at the outer region in current numerical simulations. In order to investigate the dependence of the light curve on the mass profiles in the thin-diffusion phase, we consider other forms of the mass profile and compare the light curve with that of equation (\ref{sec2-1:rho}). We adopt an exponential profile 
\begin{eqnarray}\label{sec3:rho}
\rho(t,v)&=&\rho_0\left(\frac{t}{t_0}\right)^{-3}\left(\frac{v}{v_{\rm min}}\right)^{-\beta} \nonumber \\
& & \times\exp\left(-\frac{v-0.5v_{\max}}{v_{\max}'-v}\right).
\end{eqnarray}
We introduce an additional free parameter $v_{\max}'(\ge v_{\max})$ and the ejecta expand $v_{\min}\le v\le v_{\max}'$. The calculations are the same except for using $\Delta r\sim v_{\max}'t - r_{\rm diff}$. We fix the mass with velocity larger than $0.5v_{\max}$ and calculate three models for $v_{\max}'=0.4c, 0.5c, 0.6c$. For the other parameters, we adopt from the fiducial parameter set in table 1. We show the results in the $r$-process model in figure \ref{figure:exp_comp}. Since we fix the mass with velocity larger than $0.5v_{\max}$, the bolometric luminosities at the time $t\sim t_{\times}$ are almost the same values. In the thin-diffusion phase ($t\ll t_{\times}$), both the luminosity and the temperature are smaller than the fiducial model. This is because the density at the front of the ejecta is reduced in this mass profile. Since the maximum velocity effectively becomes large and the adiabatic cooling becomes efficient, these effects for luminosity and temperature should be also seen in the engine model. We conclude that the luminosities in the thin-diffusion phase ($t\ll t_{\times}$) have uncertainties at least with $\sim1-2$ mag, which originates from the uncertainty of the outermost mass profile.

Note that the emission from the ejecta with the mass profile discussed here reduces the tension between the light curve in the $r$-process model and the upper limits of the deep optical observations (GRB 050509 and GRB 080905A) as discussed in section \ref{other}. Especially, the optical luminosity in the case $v_{\max}'=0.6c$ (thin-dot-dashed line) significantly decreases after $t\gtrsim10^5$ s. Therefore, the ejecta with relatively shallow mass distribution at the front of the ejecta is able to explain the current optical follow-up observations in the $r$-process model.

\subsection{Implications to Discriminate Two Models}
\label{implication}

In the fiducial parameter set, the light curves for two models in the optical and infrared bands are similar (figure \ref{figure:observation}). In figure \ref{figure:parameter}, the allowed parameter region to explain the observation of GRB 130603B for both models are also overlapped. Recall that emission from ejecta is described as blackbody radiation for the two models, whose spectrum is narrow in bands. Therefore, no excess at other wavelengths is expected and also the prediction is consistent with radio observations \citep{Fong+14}. Therefore, it is difficult to discriminate two models from the currently available observational data. 

In the $r$-process model, the light curve with mass profile $\rho\propto v^{-\beta}$ ($3\lesssim\beta\lesssim4$) and a parameter set which explains the infrared excess detected from GRB 130603B cannot explain the upper limits obtained from the deep optical observations of some short GRBs. Therefore, if both stringent optical upper limits at $\sim10^5$ s and bright infrared emission at $\sim10^6$ s are simultaneously obtained from a single event (with a difference larger than two magnitudes $M_{\rm optical}(\sim10^5~{\rm s}) - M_{\rm infrared}(\sim10^6~{\rm s})\gtrsim2$ mag ), the $r$-process model is significantly restricted. For the engine model, these observations give a constraint for the temperature distribution in the ejecta, which may give new insights into the activity of the central engine. 

As shown in the middle panel of figure \ref{figure:evolution}, the bolometric luminosity in the $r$-process model from the whole ejecta (blue long-dashed line), including low temperature and/or X-rays and $\gamma$-rays produced directly in radioactive decays \citep{Chu14}, declines more gradually than that for the engine model. This is because there is no energy injection after the time $t>t_{\rm inj}$ for the engine model. Then, the luminosity significantly decreases when photons at the inner edge of the ejecta begin to diffuse out (see the middle panel of figure \ref{figure:evolution} and figure \ref{figure:observation}). The luminosity from the whole ejecta can be described as $L_{\rm bol}\sim M_{\rm ej}\dot{\epsilon}$ in the transparent phase. The index of time $t$ is determined by the nuclear heating rate, $\alpha\sim1.3$. Therefore, the two models are distinguishable by observing the temporal evolution of bolometric luminosity from the whole ejecta in this phase. 

\section{SUMMARY}
\label{summary}

We calculated the light curves of macronovae by developing analytical models. We modeled the ejecta based on the results of numerical simulations for a merger of binary NSs. In addition to the nuclear decay of $r$-process elements (the $r$-process model which is often discussed), we considered another heating mechanism for the ejecta, the engine-driven shock (engine model). We compared the results with the optical and infrared observations of the first macronova candidate associated with GRB 130603B, and showed that both models can explain the observations. In order to reproduce the observed light curve, the $r$-process model requires relatively large ejecta mass $M_{\rm ej}\gtrsim0.07M_{\odot}$ which is mainly determined by the observed infrared luminosity $\sim10^{41}$ erg s$^{-1}$ at $\sim10^6$ s. In the engine model, the internal energy of ejecta, which mainly determines the observed luminosity, does not depend on the ejecta mass. Then, unless the entire of the ejecta is effectively thin (the diffusion time is smaller than the dynamical time, $t_{\rm diff}<t$, at the inner edge of the ejecta) \footnote{Since there is no energy injection after $t>t_{\rm inj}\sim10^2$ s in the engine model, the luminosity rapidly decreases after the time $t_{\rm tr}\propto M_{\rm ej}^{1/2}$ when photons can diffuse out from the inner edge of the ejecta (section \ref{transparent}).}, the required ejecta mass is $M_{\rm ej}\gtrsim0.02M_{\odot}$, which is comparable to the recent numerical simulation results. The initial internal energy $E_{\rm int0}$ and the injection time $t_{\rm inj}$ are required as $(E_{\rm int0}/10^{51}{\rm erg})(t_{\rm inj}/10^2{\rm s})\gtrsim 1$, which is consistent with the observed extended emission of short GRBs, $E_{\rm iso}\sim10^{50}-10^{51}$ erg and $t_{\rm dur}\sim10-10^2$ s. The required minimum velocity is about $v_{\min}\gtrsim0.05c$ for both models, which is mainly determined by the constraint for the observed temperature $\lesssim4\times10^3$ K at $\sim10^6$ s. For the range of the opacity $\kappa\lesssim3$ cm$^2$ g$^{-1}$
, it is difficult for both models to explain the observations of macronova associated with GRB 130603B by the ejecta mass less than $M_{\rm ej}<0.2M_{\odot}$. 

If macronovae are identical, the upper limits on the luminosity obtained in the deep optical observations of other short GRBs give stringent constraints on the $r$-process model. On the other hand, the engine model satisfies these constraints if the temperature profile is centrally concentrated in the ejecta (large $\xi$). Thus, if the difference between the optical magnitude at $\sim10^5$ s and the infrared magnitude at $\sim10^6$ s is larger than $\sim2$ mag in a single event, the $r$-process model is difficult to explain the observations unless the front of the ejecta has much shallow mass distribution. Another difference in the light curves between two models is the bolometric luminosity at the transparent phase when dynamical time is smaller than the diffusion time at the inner edge of the ejecta $r_{\rm in}$. Although the optical and infrared luminosities rapidly decrease in the transparent phase, the bolometric luminosity from the whole ejecta, including lower frequency than near-infrared band and/or X-rays and $\gamma$-rays produced directly in radioactive decays, is determined by the energy injection rate of nuclear decay, $\dot{\epsilon}\propto t^{-\alpha}~(\alpha\sim1.3)$. For the engine model, the bolometric luminosity decreases rapidly in this phase (faster than $t^{-2}$). Therefore, we expect that the light curve of the bolometric luminosity from the whole ejecta can distinguish between two heating mechanisms. 

Our results show that early light curves depend on the density profile of the outermost edge of the ejecta. It is necessary to develop a method to calculate the low-density region of the ejecta in either the analytical or numerical ways in order to precisely predict the early light curves of macronovae. 

\acknowledgments
We thank K. Asano, K. Kashiyama, K. Kiuchi, H. Nagakura, T. Nakamura, Y. Sekiguchi, M. Shibata for fruitful discussions. This work is supported by KAKENHI 24103006 (S.K., K.I.), 24.9375 (H.T.), 24000004, 26247042, 26287051 (K.I.). 

\renewcommand{\theequation}{A-\arabic{equation}}
\setcounter{equation}{0}
\section*{APPENDIX A. Analytic formulae for macronova light curves}

We summarize the formula for the observed temperature and bolometric luminosity. The detailed derivation of equations in this section is described in sections 2 and 3. 

Since we assume that the observed luminosity and temperature approximate the luminosity and temperature at the diffusion radius $r_{\rm diff}$ (section 3), we need to calculate the diffusion radius. For the dynamics of the ejecta, we assume an isotropic and homologous expansion. Then, the velocity of ejecta $v$ is described by equation (\ref{sec2:v})
\begin{eqnarray}
v\sim r/t
\end{eqnarray}
where the radius $r$ originates the central engine and the time $t$ is measured from the time when a compact binary merges. As in section \ref{fiducial}, we calculate the diffusion radius $r_{\rm diff}$ from the condition that the diffusion time equals the dynamical time, $t_{\rm diff}=t$. The diffusion time is described by equation (\ref{sec2-1:t_diff}) as
\begin{eqnarray}
t_{\rm diff}\sim\tau\frac{\Delta r}{c},
\end{eqnarray}
where $c$ is the speed of the light, $\tau$ is the optical depth described by equation (\ref{sec2-1:tau}) as
\begin{eqnarray}
\tau=\left\{ \begin{array}{ll}
{\displaystyle \int_{r_{\rm diff}}^{r_{\rm out}}\kappa\rho dr} &~~(r_{\rm diff} > 0.5r_{\rm out}) \\ 
 & \\
{\displaystyle \int_{r_{\rm diff}}^{2r_{\rm diff}}\kappa\rho dr} &~~(r_{\rm diff} \le 0.5r_{\rm out} ), \\
\end{array} \right.
\end{eqnarray}
and $\Delta r$ is the width of the diffusion region described by the equation (\ref{sec2-1:Deltar}) as
\begin{eqnarray}
\Delta r\sim \left\{ \begin{array}{ll}
 {\displaystyle r_{\rm out}-r_{\rm diff}} &~~(r_{\rm diff}>0.5r_{\rm out}) \\
 & \\
{\displaystyle r_{\rm diff}} &~~(r_{\rm diff}\le 0.5r_{\rm out}). \\
\end{array} \right.
\end{eqnarray}
In the calculation of the optical depth $\tau$, we use the spatially uniform value of the optical depth $\kappa$ with grey approximation and the ejecta mass density $\rho(t,v)$ described by equation (\ref{sec2-1:rho}) as,
\begin{eqnarray}
\rho (t,v)=\rho_0 \left(\frac{t}{t_0}\right)^{-3}\left(\frac{v}{v_{\min}}\right)^{-\beta},
\end{eqnarray}
where $\rho_0$ and $t_0$ are normalized factors, and $v_{\min}$ is the velocity at the inner edge of the ejecta. The radius $r_{\rm out}$ is the outer edge of the ejecta, described by equation (\ref{sec2-1:r_out}) as
\begin{eqnarray}
r_{\rm out}=v_{\max}t,
\end{eqnarray}
where the velocity $v_{\max}$ is at the outer edge of the ejecta. The radius at the inner edge of the ejecta $r_{\rm in}$ is described by equation (\ref{sec2-1:r_in}) as
\begin{eqnarray}
r_{\rm in}=v_{\min}t.
\end{eqnarray}
The normalization factor $\rho_0t_0^3$ in the profile of the mass density is determined by the ejecta mass $M_{\rm ej}$ (in equation \ref{sec2-1:M_ej}) as
\begin{eqnarray}
M_{\rm ej}=4\pi\int_{v_{\min}t_0}^{v_{\max}t_0}\rho(t_0,v)r^2dr.
\end{eqnarray} 

We consider two heating sources of the ejecta. In the $r$-process model, the internal energy of the ejecta is determined by the nuclear heating rate of the $r$-process element described in equation (\ref{sec2-1:dotepsilon}) as
\begin{eqnarray}
\dot{\epsilon}=\dot{\epsilon}_0\left(\frac{t}{1{\rm day}}\right)^{-\alpha}.
\end{eqnarray}
For the engine model, we assume the temperature profile of the ejecta $T(t,v)$ as a result of the activity of the central engine, described in equation (\ref{sec2-1:T}) as
\begin{eqnarray}
T(t,v)\sim T_0\left(\frac{t}{t_{\rm inj}}\right)^{-1}\left(\frac{v}{v_{\min}}\right)^{-\xi},
\end{eqnarray}
where $T_0$ is the normalization factor. This factor is determined by injected internal energy $E_{\rm int0}$ at the time $t_{\rm inj}$, which is described in equation (\ref{sec2-1:E_int}) as
\begin{eqnarray}
E_{\rm int0}=4\pi\int_{v_{\min}t_{\rm inj}}^{v_{\max}t_{\rm inj}}aT^4(t_{\rm inj},v)r^2dr.
\end{eqnarray}

Observed temperatures in the $r$-process model and the engine model are given by
\begin{eqnarray}
T_{\rm obs}\sim\left\{ \begin{array}{l}
{\displaystyle \left[\frac{\dot{\epsilon}t\rho(t,v_{\rm diff})}{a}\right]^{1/4}} \\
~~~~~~~~~~~~~~~(r_{\rm diff}>r_{\rm in}) \\
  \\
{\displaystyle \left[\frac{\dot{\epsilon}t\rho(t,v_{\min})}{a}\right]^{1/4}} \\
~~~~~~~~~~~~~~~(0.5r_{\rm in} < r_{\rm diff} \le r_{\rm in}) \\
  \\
{\displaystyle 0} \\
~~~~~~~~~~~~~~~(r_{\rm diff} \le 0.5r_{\rm in}), \\
\end{array} \right.
\end{eqnarray}
and by
\begin{eqnarray}
T_{\rm obs}\sim \left\{ \begin{array}{l}
{\displaystyle T_0\left(\frac{t}{t_{\rm inj}}\right)^{-1}\left(\frac{v_{\rm diff}}{v_{\min}}\right)^{-\xi}} \\
~~~~~~~~~~~~~~~~~(r_{\rm diff} > r_{\rm in}) \\
 \\
{\displaystyle T_0\left(\frac{t}{t_{\rm inj}}\right)^{-1}} \\
~~~~~~~~~~~~~~~~~(0.5r_{\rm in} < r_{\rm diff} \le r_{\rm in}) \\
 \\
0 \\
~~~~~~~~~~~~~~~~~(r_{\rm diff} \le 0.5r_{\rm in}), \\
\end{array} \right.
\end{eqnarray}
respectively. Note that we do not use the approximation $\rho(v,t)\sim\rho(v_{\max},t)$ and $T(v,t)\sim T(v_{\max},t)$ in the thin-diffusion case. The bolometric luminosities for the $r$-process model and the engine model are given by
\begin{eqnarray}
L_{\rm bol}\sim\left\{ \begin{array}{l}
{\displaystyle 4\pi \int_{r_{\rm diff}}^{r_{\rm out}}\rho(v,t)\dot{\epsilon}r^2dr} \\
~~~~~~~~~~~~~~~(r_{\rm diff} > 0.5r_{\rm out}) \\
 \\
{\displaystyle 4\pi \int_{r_{\rm diff}}^{2r_{\rm diff}}\rho(v,t)\dot{\epsilon}r^2dr} \\
~~~~~~~~~~~~~~~(r_{\rm in} < r_{\rm diff} \le 0.5r_{\rm out}) \\
 \\
{\displaystyle 4\pi \int_{r_{\rm in}}^{2r_{\rm diff}}\rho(v,t)\dot{\epsilon}r^2dr} \\
~~~~~~~~~~~~~~~(0.5r_{\rm in} < r_{\rm diff} \le r_{\rm in}) \\
 \\
{\displaystyle 0} \\
~~~~~~~~~~~~~~~(r_{\rm diff} \le 0.5r_{\rm in}), \\
\end{array} \right.
\end{eqnarray}
and
\begin{eqnarray}
L_{\rm bol}\sim\left\{ \begin{array}{l}
{\displaystyle 4\pi\int_{r_{\rm diff}}^{r_{\rm out}}\frac{aT_{\rm obs}^4}{t}r^2dr} \\
~~~~~~~~~~~~~~~(r_{\rm diff} > 0.5r_{\rm out}) \\
 \\
{\displaystyle 4\pi\int_{r_{\rm diff}}^{2r_{\rm diff}}\frac{aT_{\rm obs}^4}{t}r^2dr} \\
~~~~~~~~~~~~~~~(r_{\rm in} < r_{\rm diff} \le 0.5r_{\rm out}) \\
 \\
{\displaystyle 4\pi\int_{r_{\rm in}}^{2r_{\rm diff}}\frac{aT_{\rm obs}^4}{t}r^2dr} \\
~~~~~~~~~~~~~~~(0.5r_{\rm in} < r_{\rm diff} \le r_{\rm in}) \\
 \\
0 \\
~~~~~~~~~~~~~~~(r_{\rm diff} \le 0.5r_{\rm in}), \\
\end{array} \right.
\end{eqnarray}
respectively. An example of the calculated result is shown in figure \ref{figure:evolution}. 

We present the numerical values with the parameter dependence for later use. Unlike equations (\ref{sec2-2-2:t_times}) and (\ref{sec2-2-3:t_tr}), we include the contribution from subdominant terms to the numerical values when we integrate equations. Some of the subdominant terms include the ratio $v_{\max}/v_{\min}$. Hereafter, the value $v_{\max}/v_{\min}=4$ in subdominant terms are fixed and are not included in the parameter dependence. We introduce the normalized quantities $M_{\rm ej,0.1}\equiv M_{\rm ej}/0.1M_{\odot}$, $v_{\min, 0.1}\equiv v_{\min}/0.1c$, $v_{\max,0.4}\equiv v_{\max}/0.4c$, $\kappa_{10}\equiv\kappa/10$ cm$^2$ g$^{-1}$, $E_{\rm int0,51}\equiv E_{\rm int0}/10^{51}$ erg and $t_{\rm inj,2}\equiv t_{\rm inj}/10^2$ s. For other parameters, we fix the index of the mass density profile $\beta=3.5$ and the parameters of the nuclear heating rate $\dot{\epsilon}_0=2\times10^{10}$ erg s$^{-1}$ g$^{-1}$ and $\alpha=1.3$. We also introduce the normalized time $t_5\equiv t/10^5$ s and $t_6\equiv t/10^6$ s. The values of observed temperature and bolometric luminosity in the thin-diffusion phase are
\begin{eqnarray}\label{app:T_obs,thin}
T_{\rm obs}\sim\left\{ \begin{array}{l}
{\displaystyle 5.63\times10^3~~{\rm K}} \\
~~{\displaystyle \times M_{\rm ej,0.1}^{0.25}v_{\min,0.1}^{0.125}v_{\max,0.4}^{-0.875}t_5^{-0.825}}  \\
~~~~~~~~~~~~~~~~~~~~~~~~~~~(r-{\rm process}) \\
 \\
{\displaystyle 6.72\times10^3~~{\rm K}}  \\
~~{\displaystyle \times E_{\rm int0,51}^{0.25}t_{\rm inj,2}^{0.25}v_{\min,0.1}^{0.25}v_{\max,0.4}^{-1}t_5^{-1}} \\
~~~~~~~~~~~~~~~~~~~~~~~~~~~({\rm engine},~\xi=1) \\
 \\
{\displaystyle 2.34\times10^3~~{\rm K}}  \\
~~{\displaystyle \times E_{\rm int0,51}^{0.25}t_{\rm inj,2}^{0.25}v_{\min,0.1}^{1.25}v_{\max,0.4}^{-2}t_5^{-1}} \\
~~~~~~~~~~~~~~~~~~~~~~~~~~~({\rm engine},~\xi=2) \\
 \\
{\displaystyle 6.77\times10^2~~{\rm K}}  \\
~~{\displaystyle \times E_{\rm int0,51}^{0.25}t_{\rm inj,2}^{0.25}v_{\min,0.1}^{2.25}v_{\max,0.4}^{-3}t_5^{-1}} \\
~~~~~~~~~~~~~~~~~~~~~~~~~~~({\rm engine},~\xi=3), \\
\end{array} \right.
\end{eqnarray}
and
\begin{eqnarray}\label{app:L_bol,thin}
L_{\rm bol}\sim\left\{ \begin{array}{l}
{\displaystyle 3.51\times10^{41}~~{\rm erg~s}^{-1}} \\
~~{\displaystyle \times \kappa_{10}^{-0.5}M_{\rm ej,0.1}^{0.5}v_{\min,0.1}^{0.25}v_{\max,0.4}^{0.25}t_5^{-0.3}} \\
~~~~~~~~~~~~~~~~~~~~~~~~~~~~~~(r-{\rm process}) \\
 \\
{\displaystyle 7.10\times10^{41}~~{\rm erg~s}^{-1}} \\
~~{\displaystyle \times \kappa_{10}^{-0.5}M_{\rm ej,0.1}^{-0.5}E_{\rm int0,51}t_{\rm inj,2}v_{\min,0.1}^{0.75}v_{\max,0.4}^{-0.25}t_5^{-1}} \\
~~~~~~~~~~~~~~~~~~~~~~~~~~~~~~({\rm engine},~\xi=1) \\
 \\
{\displaystyle 1.04\times10^{40}~~{\rm erg~s}^{-1}} \\
~~{\displaystyle \times \kappa_{10}^{-0.5}M_{\rm ej,0.1}^{-0.5}E_{\rm int0,51}t_{\rm inj,2}v_{\min,0.1}^{4.75}v_{\max,0.4}^{-4.25}t_5^{-1}} \\
~~~~~~~~~~~~~~~~~~~~~~~~~~~~~~({\rm engine},~\xi=2) \\
 \\
{\displaystyle 7.32\times10^{37}~~{\rm erg~s}^{-1}} \\
~~{\displaystyle \times \kappa_{10}^{-0.5}M_{\rm ej,0.1}^{-0.5}E_{\rm int0,51}t_{\rm inj,2}v_{\min,0.1}^{8.75}v_{\max,0.4}^{-8.25}t_5^{-1}} \\
~~~~~~~~~~~~~~~~~~~~~~~~~~~~~~({\rm engine},~\xi=3), \\
\end{array} \right.
\end{eqnarray}
respectively. 

The transition time from the thin-diffusion phase to the thick-diffusion phase $t_{\times}$ is
\begin{eqnarray}\label{app:t_times}
t_{\times}&=&\sqrt{\frac{2^{\beta-4}(\beta-3)(1-2^{1-\beta})\kappa M_{\rm ej}}{\pi(\beta-1)[1-(v_{\max}/v_{\min})^{3-\beta}]cv_{\max}}\left(\frac{v_{\max}}{v_{\min}}\right)^{3-\beta}} \nonumber \\
 \nonumber \\
&\sim&4.53\times10^5~\kappa_{10}^{0.5}M_{\rm ej,0.1}^{0.5}v_{\min,0.1}^{0.25}v_{\max,0.4}^{-0.75}~{\rm s}.
\end{eqnarray}
The values of the observed temperature and bolometric luminosity in the thick-diffusion phase are
\begin{eqnarray}
T_{\rm obs}\sim\left\{ \begin{array}{l}
{\displaystyle 3.89\times10^3~~{\rm K}} \\
~~{\displaystyle \times \kappa_{10}^{-0.583}M_{\rm ej,0.1}^{-0.333}v_{\min,0.1}^{-0.167}t_6^{0.342}} \\
~~~~~~~~~~~~~~~~~~~~~~~~~~~~~~(r-{\rm process}) \\
 \\
{\displaystyle 3.86\times10^3~~{\rm K}} \\
~~{\displaystyle \times \kappa_{10}^{-0.667}M_{\rm ej,0.1}^{-0.667}E_{\rm int0,51}^{0.25}t_{\rm inj,2}^{0.25}v_{\min,0.1}^{-0.083}t_6^{0.333}} \\
~~~~~~~~~~~~~~~~~~~~~~~~~~~~~~({\rm engine},~\xi=1) \\
 \\
{\displaystyle 7.73\times10^3~~{\rm K}} \\
~~{\displaystyle \times \kappa_{10}^{-1.333}M_{\rm ej,0.1}^{-1.333}E_{\rm int0,51}^{0.25}t_{\rm inj,2}^{0.25}v_{\min,0.1}^{0.583}t_6^{1.667}} \\
~~~~~~~~~~~~~~~~~~~~~~~~~~~~~~({\rm engine},~\xi=2) \\
 \\
{\displaystyle 1.29\times10^4~~{\rm K}} \\
~~{\displaystyle \times \kappa_{10}^{-2}M_{\rm ej,0.1}^{-2}E_{\rm int0,51}^{0.25}t_{\rm inj,2}^{0.25}v_{\min,0.1}^{1.25}t_6^3} \\
~~~~~~~~~~~~~~~~~~~~~~~~~~~~~~({\rm engine},~\xi=3), \\
\end{array} \right.
\end{eqnarray}
and
\begin{eqnarray}
L_{\rm bol}\sim\left\{ \begin{array}{l}
{\displaystyle 1.16\times10^{41}~~{\rm erg~s}^{-1}} \\
{\displaystyle ~~\times \kappa_{10}^{-0.333}M_{\rm ej,0.1}^{0.667}v_{\min,0.1}^{0.333}t_6^{-0.633}} \\
{\displaystyle ~~~~~~~~~~~~~~~~~~~~~~~~~~~~~~(r-{\rm process})} \\
 \\
{\displaystyle 9.59\times10^{40}~~{\rm erg~s}^{-1}} \\
{\displaystyle ~~\times \kappa_{10}^{-0.667}M_{\rm ej,0.1}^{-0.667}E_{\rm int0,51}t_{\rm inj,2}v_{\min,0.1}^{0.667}t_6^{-0.667}} \\
{\displaystyle ~~~~~~~~~~~~~~~~~~~~~~~~~~~~~~({\rm engine},~\xi=1)} \\
 \\
{\displaystyle 5.96\times10^{41}~~{\rm erg~s}^{-1}} \\
{\displaystyle ~~\times \kappa_{10}^{-3.333}M_{\rm ej,0.1}^{-3.333}E_{\rm int0,51}t_{\rm inj,2}v_{\min,0.1}^{3.333}t_6^{4.667}} \\
{\displaystyle ~~~~~~~~~~~~~~~~~~~~~~~~~~~~~~({\rm engine},~\xi=2)} \\
 \\
{\displaystyle 2.62\times10^{42}~~{\rm erg~s}^{-1}} \\
{\displaystyle ~~\times \kappa_{10}^{-6}M_{\rm ej,0.1}^{-6}E_{\rm int0,51}t_{\rm inj,2}v_{\min,0.1}^6t_6^{10}} \\
{\displaystyle ~~~~~~~~~~~~~~~~~~~~~~~~~~~~~~({\rm engine},~\xi=3),} \\
\end{array} \right.
\end{eqnarray}
respectively. Note that since the diffusion radius $r_{\rm diff}$ cannot be analytically described in the thin-diffusion phase, we use the approximations $\rho(v,t)\sim\rho(v_{\max},t)$ and $T(v,t)\sim T(v_{\max},t)$ in equations (\ref{app:T_obs,thin}) and (\ref{app:L_bol,thin}). These approximations make discontinuity at the transition time $t_{\times}$. The ratios of the temperature in the thick-diffusion phase to the temperature in the thin-diffusion phase for the $r$-process $A_{\rm T,r}$ and the engine model $A_{\rm T,e}$  at the time $t_{\times}$ are
\begin{eqnarray}
A_{\rm T,r}&=&2^{\beta/4} \nonumber \\
&\sim&1.83
\end{eqnarray}
and
\begin{eqnarray}
A_{\rm T,e}&=&2^{\xi} \nonumber \\
&\sim&\left\{ \begin{array}{ll}
 2.00 &~~(\xi=1) \\
 4.00 &~~(\xi=2) \\
 8.00 &~~(\xi=3), \\
\end{array} \right.
\end{eqnarray}
respectively. The ratios of the luminosity in the thick-diffusion phase to the luminosity in the thin-diffusion phase for the $r$-process model $A_{\rm L,r}$ and the engine model $A_{\rm L,e}$ at the time $t_{\times}$ are
\begin{eqnarray}
A_{\rm L,r}&=&2^{\frac{\beta-4}{2}}\left(\frac{1-2^{3-\beta}}{\beta-3}\right)\sqrt{\frac{\beta-1}{1-2^{1-\beta}}} \nonumber \\
&\sim&0.858
\end{eqnarray}
and
\begin{eqnarray}
A_{\rm L,e}&=&2^{\frac{-\beta-4+8\xi}{2}}\left(\frac{1-2^{3-4\xi}}{4\xi-3}\right)\sqrt{\frac{\beta-1}{1-2^{1-\beta}}} \nonumber \\
&\sim&\left\{ \begin{array}{ll}
 1.04 &~~(\xi=1) \\
 6.42 &~~(\xi=2) \\
 58.8 &~~(\xi=3), \\
\end{array} \right.
\end{eqnarray}
respectively.

The transition time from the thick-diffusion phase to the transparent phase $t_{\rm tr}$ is
\begin{eqnarray}\label{app:t_tr}
t_{\rm tr}&=&\sqrt{\frac{(\beta-3)(1-2^{1-\beta})\kappa M_{\rm ej}}{4\pi(\beta-1)[1-(v_{\max}/v_{\min})^{3-\beta}]cv_{\min}}} \nonumber \\
 \nonumber \\ 
&\sim&7.62\times10^5~\kappa_{10}^{0.5}M_{\rm ej,0.1}^{0.5}v_{\min,0.1}^{-0.5}~{\rm s}.
\end{eqnarray}
We introduce another transition time $t_{\rm tr2}$ when the upper limit of the integral for the luminosity $2r_{\rm diff}$ reaches the inner edge of the ejecta $r_{\rm in}$, 
\begin{eqnarray}
t_{\rm tr2}&=&2^{\frac{\beta-2}{2}}t_{\rm tr} \nonumber \\
&\sim&1.28\times10^6~\kappa_{10}^{0.5}M_{\rm ej,0.1}^{0.5}v_{\min,0.1}^{-0.5}~{\rm s}.
\end{eqnarray}
The values of the observed temperature and bolometric luminosity in the transparent phase ($t_{\rm tr}\le t<t_{\rm tr2}$) are
\begin{eqnarray}
T_{\rm obs}\sim\left\{ \begin{array}{l}
{\displaystyle 2.83\times10^3~~{\rm K}} \\
~~{\displaystyle \times M_{\rm ej,0.1}^{0.25}v_{\min,0.1}^{-0.75}t_6^{-0.825}} \\
~~~~~~~~~~~~~~~~~~~~(r-{\rm process}) \\
 \\
{\displaystyle 2.69\times10^3~~{\rm K}} \\
~~{\displaystyle \times E_{\rm int0,51}^{0.25}t_{\rm inj,2}^{0.25}v_{\min,0.1}^{-0.75}t_6^{-1}} \\
~~~~~~~~~~~~~~~~~~~~({\rm engine},~\xi=1) \\
 \\
{\displaystyle 3.74\times10^3~~{\rm K}} \\
~~{\displaystyle \times E_{\rm int0,51}^{0.25}t_{\rm inj,2}^{0.25}v_{\min,0.1}^{-0.75}t_6^{-1}} \\
~~~~~~~~~~~~~~~~~~~~({\rm engine},~\xi=2) \\
 \\
{\displaystyle 4.33\times10^3~~{\rm K}} \\
~~{\displaystyle \times E_{\rm int0,51}^{0.25}t_{\rm inj,2}^{0.25}v_{\min,0.1}^{-0.75}t_6^{-1}} \\
~~~~~~~~~~~~~~~~~~~~({\rm engine},~\xi=3) \\
\end{array} \right.
\end{eqnarray}
and 
\begin{eqnarray}
L_{\rm bol}\sim\left\{ \begin{array}{l}
{\displaystyle 3.30\times10^{41}~~{\rm erg~s}^{-1}} \\
~~{\displaystyle \times M_{\rm ej,0.1}t_6^{-1.3}\left[1-\left(\frac{t}{t_{\rm tr2}}\right)^{0.667}\right]} \\
~~~~~~~~~~~~~~~~~~~~~~~~~~~~~~(r-{\rm process}) \\
  \\
{\displaystyle 1.33\times10^{41}~~{\rm erg~s}^{-1}}  \\
~~{\displaystyle \times E_{\rm int0,51}t_{\rm inj,2}t_6^{-2}\left[1-\left(\frac{t}{t_{\rm tr2}}\right)^{1.333}\right]} \\
~~~~~~~~~~~~~~~~~~~~~~~~~~~~~~({\rm engine},~\xi=1), \\
  \\
{\displaystyle 1.00\times10^{41}~~{\rm erg~s}^{-1}}  \\
~~{\displaystyle \times E_{\rm int0,51}t_{\rm inj,2}t_6^{-2}\left[1-\left(\frac{t}{t_{\rm tr2}}\right)^{6.667}\right]} \\
~~~~~~~~~~~~~~~~~~~~~~~~~~~~~~({\rm engine},~\xi=2), \\
  \\
{\displaystyle 1.00\times10^{41}~~{\rm erg~s}^{-1}}  \\
~~{\displaystyle \times E_{\rm int0,51}t_{\rm inj,2}t_6^{-2}\left[1-\left(\frac{t}{t_{\rm tr2}}\right)^{12}\right]} \\
~~~~~~~~~~~~~~~~~~~~~~~~~~~~~~({\rm engine},~\xi=3), \\
\end{array} \right.
\end{eqnarray}
respectively.



\end{document}